\documentclass[10pt,letterpaper]{article}
\usepackage[top=0.85in,left=1.0in,footskip=0.75in,marginparwidth=2in]{geometry}

\usepackage[utf8]{inputenc}
\usepackage{cite}
\usepackage{nameref,hyperref}
\usepackage{microtype}
\DisableLigatures[f]{encoding = *, family = * }

\raggedright
\usepackage{changepage}
\usepackage[aboveskip=1pt,labelfont=bf,labelsep=period,singlelinecheck=off]{caption}

\usepackage{color}
\definecolor{Gray}{gray}{.25}
\usepackage{graphicx}
\usepackage{sidecap}

\usepackage{wrapfig}
\usepackage[pscoord]{eso-pic}
\usepackage[fulladjust]{marginnote}
\usepackage{graphicx}
\usepackage[section]{algorithm}
\usepackage{algorithmic}
\usepackage{threeparttable}
\usepackage{bm}
\usepackage{color}
\usepackage{ragged2e}
\usepackage{subfigure}
\usepackage{mathrsfs}
\usepackage{enumitem}
\usepackage{array}
\usepackage{url}
\include{multicol}
\newcommand{\PreserveBackslash}[1]{\let\temp=\\#1\let\\=\temp}
\newcolumntype{C}[1]{>{\PreserveBackslash\centering}p{#1}}
\newcolumntype{R}[1]{>{\PreserveBackslash\raggedleft}p{#1}}
\newcolumntype{L}[1]{>{\PreserveBackslash\raggedright}p{#1}}

\renewcommand{\raggedright}{\leftskip=0pt \rightskip=0pt plus 0cm}

\makeatletter
\renewcommand{\@biblabel}[1]{\quad#1.}
\makeatother

\usepackage{lastpage,fancyhdr,graphicx}
\usepackage{epstopdf}
\fancyheadoffset[L]{2.25in}
\fancyfootoffset[L]{2.25in}

\reversemarginpar

\begin{document}
\vspace*{0.35in}

\begin{flushleft}
{\Large{\textbf{A Survey on Applications of Artificial Intelligence in Fighting Against COVID-19}}}
\newline
\\
Jianguo~Chen\textsuperscript{1,2},
Kenli~Li \textsuperscript{1},
Zhaolei Zhang\textsuperscript{2},
Keqin Li\textsuperscript{1,3},
Philip S. Yu\textsuperscript{4}
\\
\bigskip
$^{1}$ College of Computer Science and Electronic Engineering, Hunan University, Changsha, Hunan, 410082, China.
\\
$^{2}$ Donnelly Centre for Cellular and Biomolecular Research and Department of Computer Science, University of Toronto, Toronto, ON, 710049, Canada.
\\
$^{3}$ Department of Computer Science, State University of New York, New Paltz, NY, 12561, USA.
\\
$^{4}$ Department of Computer Science, University of Illinois at Chicago, Chicago, IL, 60607, USA.
\bigskip
\\
* Correspinding authors: Kenli Li (lkl@hnu.edu.cn) and Keqin Li (lik@newpaltz.edu).

\end{flushleft}

\section*{Abstract}
The COVID-19 pandemic caused by the SARS-CoV-2 virus has spread rapidly worldwide, leading to a global outbreak.
Most governments, enterprises, and scientific research institutions are participating in the COVID-19 struggle to curb the spread of the pandemic.
As a powerful tool against COVID-19, artificial intelligence (AI) technologies are widely used in combating this pandemic.
In this survey, we investigate the main scope and contributions of AI in combating COVID-19 from the aspects of disease detection and diagnosis, virology and pathogenesis, drug and vaccine development, and epidemic and transmission prediction.
In addition, we summarize the available data and resources that can be used for AI-based COVID-19 research.
Finally, the main challenges and potential directions of AI in fighting against COVID-19 are discussed.
Currently, AI mainly focuses on medical image inspection, genomics, drug development, and transmission prediction, and thus AI still has great potential in this field.
This survey presents medical and AI researchers with a comprehensive view of the existing and potential applications of AI technology in combating COVID-19 with the goal of inspiring researchers to continue to maximize the advantages of AI and big data to fight COVID-19.

\section{Introduction}
Severe Acute Respiratory Syndrome Corona-Virus 2 (SARS-CoV-2) is an emerging human infectious coronavirus.
Since December 2019, the coronavirus disease 2019 (COVID-19) caused by SARS-CoV-2 was first reported in China and later in most countries in the world \cite{wu2020new, arabi2020covid, ghinai2020first}.
The World Health Organization (WHO) announced on January 30, 2020, that the outbreak was a Public Health Emergency of International Concern (PHEIC), and confirmed COVID-19 as a pandemic on March 11, 2020.
As of February 26, 2021, this disease has been reported in 216 countries or regions around the world and has resulted in serious consequences, including 112,649,371 confirmed COVID-19 cases and 2,501,229 deaths \cite{WHO2020data}.

Most governments, enterprises, and scientific research institutions are fighting COVID-19 from all aspects to curb the spread of the disease \cite{Italy2020data, dong2020interactive, ecdc2020data, ghddi2020data}.
Various stakeholders from different institutions and backgrounds have provided abundant resources and capabilities to support this disease battle \cite{chen2020keep, ncbi2020data, gisaid2020data, nhc2020data, mediaCloud2020data, google2020data, baidu2020data}.
As far as SARS-CoV-2 is concerned, virology, origin and classification, physicochemical properties, receptor interactions, cell entry, genomic variation, and ecology are thoroughly studied \cite{andersen2020proximal, hoffmann2020sars, lu2020genomic, zhou2020pneumonia, lu2020genomic, wolfel2020virological}.
Nucleic acid testing, serologic diagnosis, and medical imaging (i.e., chest X-ray or CT imaging) are the main disease detection and diagnosis methods at present \cite{kong2020sars, chen2020clinical, huang2020clinical, to2020temporal, xu2020characteristics}.
In terms of pathogenesis, topics such as virus entry and spread, pathological findings, and immune response are the focus \cite{walls2020structure, ou2020characterization, walls2019unexpected, thevarajan2020breadth}.
In epidemiology, extensive research has been conducted on the source and spectrum of infection, clinical features, epidemiological characteristics, epidemic prediction, and transmission route tracking \cite{chan2020familial, li2020characterizing, wu2020estimating, wu2020nowcasting}.
Potential therapeutics of COVID-19 include intensive care, drug development, and vaccine development \cite{chen2020fangcang, phua2020intensive, touret2020chloroquine, liu2020hydroxychloroquine}.
In addition, communication prediction and social isolation are currently the main social control methods \cite{kandel2020health, qian2020fighting, abel2020covid, zavaleta2020covid}.

Artificial intelligence (AI) is defined as a technology that allows computers to imitate human intelligence to process things, including Machine Learning (ML), knowledge graphs, natural language processing, human-computer interaction, computer vision, biometrics, virtual reality, and augmented reality \cite{szegedy2017inception, castiglioni2020artificial, li2020artificial}.
ML can be subdivided into traditional ML and deep learning (DL).
Traditional ML methods include logistic regression, decision tree, random forest, K-nearest neighbor, Adaboost, K-means clustering, density clustering, hidden Markov models, support vector machine, Naive Bayes, etc \cite{breiman2001random, ong2020vaxign, maaten2008visualizing}.
DL is a subset of ML and is a learning method for building deep structural neural networking models \cite{greff2016lstm, skalic2019shape}.
In addition, ML techniques also include transfer learning, active learning, and evolutionary learning.
In recent years, AI technology has achieved technological breakthroughs and is widely used in various fields of intelligent medicine, including medical image inspection, disease-assisted diagnosis, surgery, hospital management, and medical big data integration \cite{szegedy2017inception}.
Moreover, AI is actively explored in emerging fields such as surgical robots, wearable devices, new drug discovery, precision medicine, epidemics prevention and control, and gene sequencing.

Encouragingly, in the short period of time since the outbreak of COVID-19, research in the fields of industry, medical, and science has successfully used advanced AI technologies in the COVID-19 battle and has achieved significant progress.
For example, AI supports the diagnosis of COVID-19 through medical image inspections and provides non-invasive detection solutions to prevent medical personnel from contracting infections \cite{castiglioni2020artificial, li2020artificial,
allam2020artificial}.
AI is used in virology research to analyze the structure of SARS-CoV-2-related proteins and predict new compounds that can be used in drug and vaccine development \cite{zhou2020network, moskal2020suggestions, bung2020novo}.
In addition, AI has achieved virus source tracking through genomic research, and has successfully discovered the relationship between SARS-CoV-2 and the bat virus, as well as the relationship between SARS-CoV and the Middle East respiratory syndrome-related coronavirus (MERS-CoV) \cite{ortega2020role, randhawa2020machine, demirci2020computational}.
Moreover, AI learns large-scale COVID-19 case data and social media data to construct epidemic transmission models to accurately predict the outbreak time, transmission route, transmission range, and impact of the disease \cite{pan2020identification, li2020preliminary, santosh2020ai, pourhomayoun2020predicting}.
AI is also widely used in epidemic prevention and social control, such as airport security inspections, patient trajectory tracking, and epidemic visualization \cite{rahmatizadeh2020role, rahmatizadeh2020role}.
There are some studies and surveys on the COVID-19 epidemic and the related machine learning or artificial intelligence applications.
For example, in \cite{shinde2020forecasting}, Shinde \emph{et al}. summarized various forecasting techniques for COVID-19, including stochastic theory, mathematical models, data science, and machine learning techniques.
In \cite{shi2020review}, Shi \emph{et al}. provided a survey of AI techniques in imaging data collection, segmentation and diagnosis of COVID-19.
However, most of the existing work only focuses on one or more narrow fields of the COVID-19 fight, leaving readers difficult to know the scope of the current research on COVID-19.

\begin{figure}[!ht]
\setlength{\abovecaptionskip}{4pt}
\setlength{\belowcaptionskip}{4pt}
  \centering
 \includegraphics[width=5.4in]{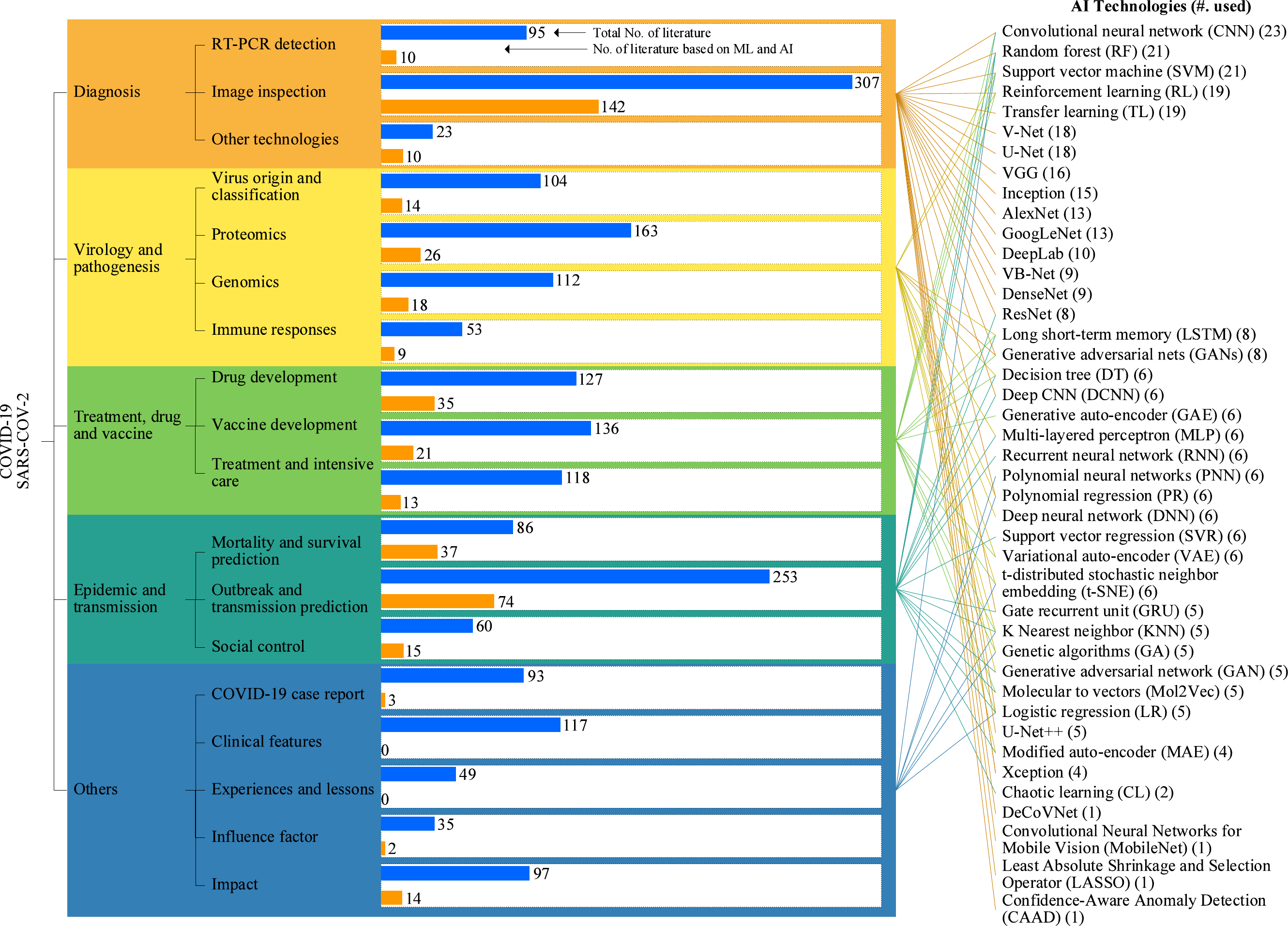}
 \caption{Main scope of AI technologies in fighting against COVID-19. We collect 2471 online publications and resources related to COVID-19, SARS-CoV-2, and 2019-nCoV from databases such as Nature, Elsevier, Google Scholar, arxiv, biorxiv, and medRxiv. Then, we filter out 443 papers that explicitly use AI methods. In addition, we list the name of AI technologies used in each field of COVID-19 research.}
\label{fig01}
\end{figure}

In this survey, we present a comprehensive view of the landscape and contributions of AI in combating COVID-19.
The main scope of AI in COVID-19 research includes disease detection and diagnosis, virology and pathogenesis, drug and vaccine development, and epidemic and transmission prediction.
Note that due to the rapid development of the COVID-19 epidemic, we cited many preprinted references for a comprehensive investigation, which still need to be assessed on their accuracy and quality through peer review.
The main scope of AI in combating COVID-19 is summarized in Fig. \ref{fig01}.

The rest of the paper is structured as follows.
Sections 2 to 5 discuss the four main scope of AI against COVID-19.
Section 6 summarizes the available data and resources to support COVID-19 research.
Section 7 highlights the challenges and potential directions in this field.
Finally, Section 8 presents the conclusion.
Table \ref{table01} gives the abbreviations and descriptions of the AI methods used in this survey.

\begin{table}[!ht]
\renewcommand{\arraystretch}{1.1}
\setlength{\abovecaptionskip}{4pt}
\setlength{\belowcaptionskip}{4pt}
\caption{Abbreviations and descriptions of the AI methods used in this survey.}
\centering
\small
\label{table01}
\tabcolsep1pt
\begin{tabular}{L{0.8in} L{1.86in} L{0.8in} L{1.86in}}
\hline
 Abbreviation       & Description        &  Abbreviation       & Description\\
\hline
CL    & Chaotic learning             & MAE & Modified auto-encoder    \\
CNN   & Convolutional neural network & MLP   & Multi-layered perceptron  \\
DCNN  & Deep CNN                     & Mol2Vec & Molecular to vectors    \\
DNN   & Deep neural network          & R-CNN   & Regional convolutional network \\
DT    & Decision tree                & PNN   & Polynomial neural networks \cite{oh2003polynomial} \\
FCN   & Fully convolutional network  & PR    & Polynomial regression \cite{de2019approximate}\\
GA    & Genetic algorithm            & RF    & Random forest \cite{breiman2001random} \\
GAE   & Generative auto-encoder      & RL    & Reinforcement learning\\
GANs  & Generative adversarial nets  & RNN   &  Recurrent neural network  \\
GAN   & Generative adversarial network & SVM & Support vector machine \\
GRU   & Gate recurrent unit          & SVR   & Support vector regression \cite{awad2015support} \\
KNN   & $K$-nearest neighbor         & TL    & Transfer learning \\
LR    & Logistic regression          & t-SNE & t-distributed stochastic neighbor embedding \cite{maaten2008visualizing}\\
LSTM  & Long short-term memory  \cite{greff2016lstm} & VAE & Variational auto-encoder \cite{skalic2019shape} \\
\hline
\end{tabular}
\end{table}


\section{Disease Detection and Diagnosis}
\label{section2}
The diagnosis of virus infection is an important part of COVID-19 research.
The current detection and diagnosis methods used for SARS-CoV-2 virus and COVID-19 disease mainly include nucleic acid testing, serological diagnosis, chest X-ray and CT image inspection, and other noninvasive methods.

\subsection{RT-PCR Detection}
Benefitting from the advantages of high sensitivity and specificity, real-time Reverse Transcriptase Polymerase Chain Reaction (RT-PCR) is the current standard detection technology in diagnosing the SARS-CoV-2 virus and bacterial infections.
Using RT-PCR, 9 RNA positives were detected from pharyngeal swabs of patients, indicating that the SARS-CoV-2 virus had spread in communities of Wuhan, China, in early January 2020 \cite{kong2020sars}.
The shedding of the SARS-CoV-2 virus detected in the throat, lungs, and feces suggests multiple routes of virus transmission \cite{wolfel2020virological, xu2020characteristics}.
However, RT-PCR faces the limitations of complicated sample preparation, low detection efficiency, and high false-negative rate \cite{kong2020sars, wang2020deep, wu2020rapid}.


Isothermal nucleic acid amplification and blood testing methods are also commonly used for rapid screening of SARS-CoV-2 \cite{kim2019simple, lu2020development, wu2020rapid}.
An ML classification method was used for blood testing to extract important routine hematological and biochemical characteristics and to provide COVID-19 classification.
In \cite{wu2020rapid}, 105 blood test reports were collected, of which 27 were positive samples from patients with confirmed COVID-19.
For comparison, negative samples were collected from patients with ordinary pneumonia, tuberculosis, and lung cancer.
Each sample contains 49 feature variables, including 24 routine hematological and 25 biochemical  parameters.
Next, the authors implemented the RF algorithm \cite{breiman2001random} on the training samples for feature learning and classification.
Based on the extracted 11 key feature variables, they built an RF classifier and tested 253 samples of 169 patients with suspected COVID-19 and obtained an accuracy of 96.97\%.
Although AI technologies rarely directly participate in RT-PCR and blood testing, the viral load and COVID-19 case data collected in these methods provide important data sources for the subsequent AI-based analysis.


\subsection{Medical Image Inspection}
Medical imaging inspection is another widely used clinical approach for COVID-19 detection and diagnosis.
COVID-19 medical image inspection mainly includes chest X-ray and lung CT imaging.
AI technology plays an important role in medical image inspection and has achieved significant results in image acquisition, organ recognition, infection region segmentation, and disease classification.
It not only greatly shortens the imaging diagnosis time of radiologists, but also improves the accuracy the diagnosis.
We will discuss in detail the contributions of AI methods to chest X-ray and lung CT imaging.

\subsubsection{CT Image Inspection}

\begin{figure}[!ht]
  \setlength{\abovecaptionskip}{4pt}
  \setlength{\belowcaptionskip}{4pt}
  \centering
  \includegraphics[width=5.45in]{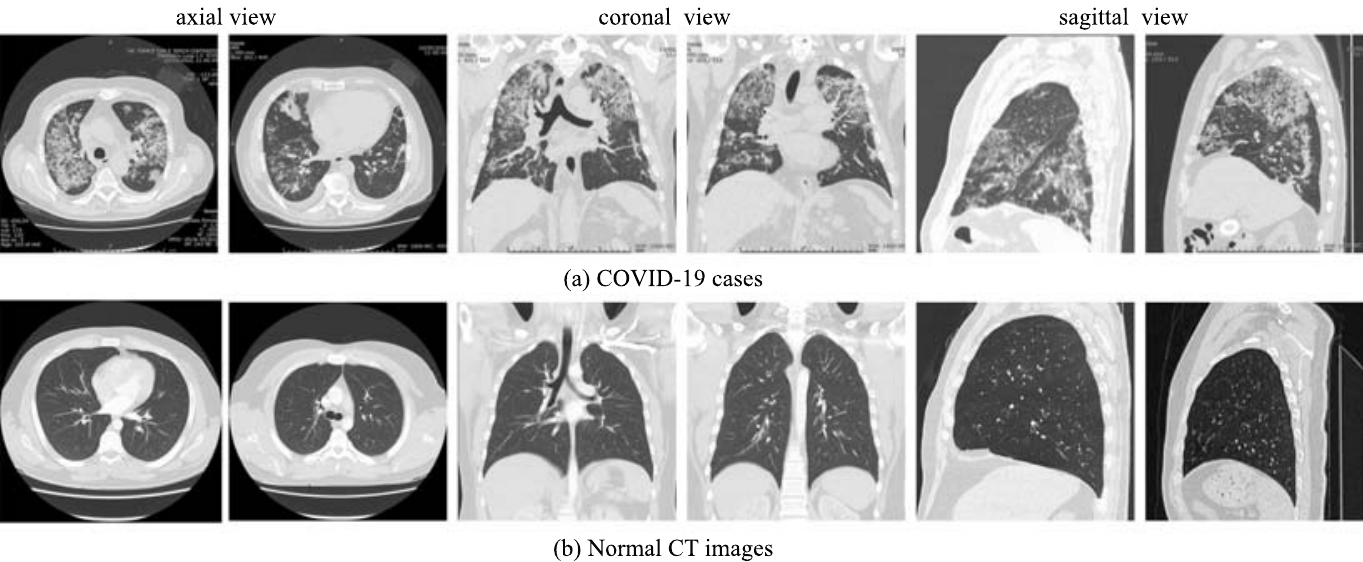}
  \caption{Examples of lung CT images of normal and COVID-19 cases.}
  \label{fig02}
\end{figure}

CT imaging provides an important basis for the early diagnosis of COVID-19.
The CT imaging manifestations of COVID-19 are mainly Ground Glass Opacity (GGO) in the periphery of the subpleural region, and some are consolidated.
If a patient's COVID-19 condition improves, the area will be absorbed and form fibrous stripes \cite{coronacases2020data, sirm2020data, radiopaedia2020data}.
Examples of lung CT images of normal and COVID-19 cases are shown in Fig. \ref{fig02}.

\begin{figure}[!ht]
\setlength{\abovecaptionskip}{4pt}
\setlength{\belowcaptionskip}{4pt}
  \centering
 \includegraphics[width=5.4in]{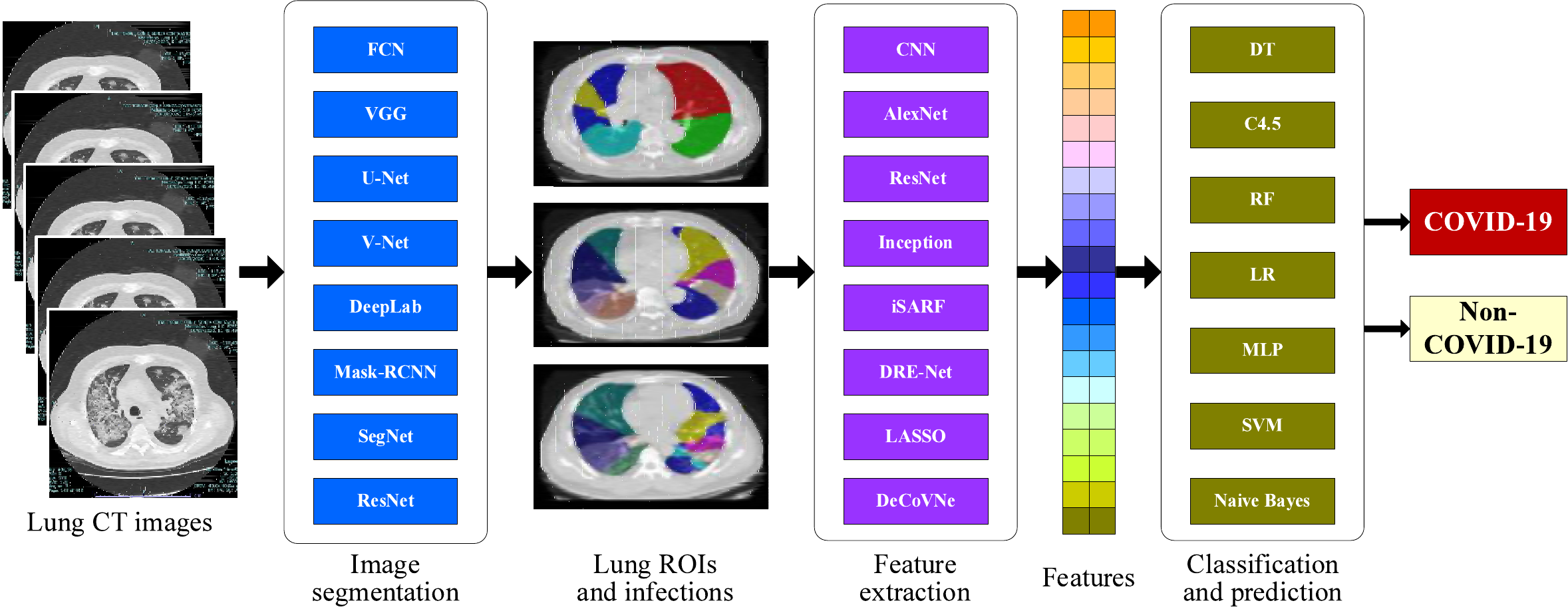}
 \caption{Representative AI architecture for CT image classification and COVID-19 inspection.}
\label{fig03}
\end{figure}

The progress of AI-based CT image inspection for COVID-19 usually includes the following steps: Region Of Interest (ROI) segmentation, lung tissue feature extraction, candidate infection region detection, and COVID-19 classification.
The representative AI architecture for CT image classification and COVID-19 inspection is shown in Fig. \ref{fig03}.

The segmentation of lung organs and ROIs is the foundational step in AI-based image inspection.
It depicts the ROIs in lung CT images (i.e., lungs, lung lobes, bronchopulmonary segments, and infected regions or lesions) for further evaluation and quantification.
Different DL models (i.e., U-Net, V-Net, and VB-Net) have been used for CT image segmentation \cite{chen2020deep, xu2020deep, li2020artificial, tang2020severity}.
In \cite{shan2020lung}, Shan \emph{et al}. collected 549 CT images from patients with confirmed COVID-19 and proposed an improved segmentation model (called VB-Net) based on the V-Net \cite{li2019volumetric} and ResNet \cite{he2016deep} models.
In \cite{chen2020deep}, Chen \emph{et al}. established a DL model based on the U-Net++ structure \cite{zhou2018unet++} to extract the ROIs from each CT image and detect the training curve of suspicious lesions.
In \cite{xu2020deep}, Xu \emph{et al}. used a 3D DL model to segment the infection regions from lung CT images.
Then, they built a classification model using ResNet and location-attention structures, and divided the segmented regional images into three categories, such as COVID-19, influenza-A viral pneumonia, and normal.
In \cite{li2020artificial}, Li \emph{et al}. used the U-Net segmentation model to extract lung organs as ROIs from each lung CT image.
In \cite{tang2020severity}, Tang \emph{et al}. used the VB-Net model \cite{shan2020lung} to accurately segment 18 lung regions and infected regions from lung CT images, and further calculated 63 quantitative features.

Focusing on the detection and location of candidate infection regions, different AI methods were proposed in \cite{gozes2020rapid, shi2020deep,wang2020deep, huang2020serial}.
In \cite{gozes2020rapid}, Gozes \emph{et al}. used commercial software to identify lung nodules and small opacity in the 3D lung volume.
Then, they constructed a DL model composed of the U-Net and ResNet structures, where the U-Net module was used to extract the ROI regions, and the ResNet model was used to detect and classify diffuse turbidity and ground glass infiltration.
In addition, they compared the CT images of 56 patients with confirmed COVID-19 and 101 non-coronavirus patients, and analyzed the CT features of COVID-19 in detail.
In \cite{shi2020deep}, Shi \emph{et al}. used a V-Net-based CNN model to segment lung organs and infected regions from lung CT images.
Then, they used the Least Absolute Shrinkage and Selection Operator (LASSO) method to calculate the best CT morphological features.
Finally, based on the best CT morphology and clinical features, the severity of COVID-19 was predicted and evaluated.
In \cite{wang2020deep}, Wang \emph{et al}. collected 195 CT images from 44 patients with COVID-19 and 258 CT images from 55 negative patients.
They used the CNN model with the Inception structure \cite{szegedy2016rethinking} to classify randomly selected ROI images and predict COVID-19 disease.
In \cite{huang2020serial}, Huang \emph{et al}. used an InferReadTM CT pneumonia tool based on AI to quantitatively evaluate changes in the lung burden of patients with COVID-19.
The tool includes three modules: lung and lobe extraction, pneumonia segmentation, and quantitative analysis.
The CT image features of COVID-19 pneumonia are divided into four types: mild, moderate, severe, and critical.

Based on ROI segmentation and candidate infection region detection, the important features of ROIs and infection regions are extracted for COVID-19 classification \cite{qi2020machine}.
In \cite{qi2020machine}, Qi \emph{et al}. collected 71 CT images from 52 patients with confirmed COVID-19 in 5 hospitals.
They used radiobiological methods to extract 1,218 features from each CT image, and then performed LR and RF methods on these features to distinguish between short-term and long-term hospital stays.
In \cite{shi2020large}, Shi \emph{et al}. used the VB-Net model \cite{shan2020lung} to segment lung and infection regions from CT images, and classified them based on 96 features (including 26 volume features, 31 digital features, 32 histogram features, and 7 surface features).
Next, they proposed an iSARF method to classify features and predict COVID-19 disease.
Comparative experiments show that the iSARF method is superior to LR, SVM, and NN methods.
In \cite{zheng2020deep}, Zheng \emph{et al}. proposed a 3D DCNN model (called DeCoVNet) to detect COVID-19 from CT images.
The proposed DeCoVNet model includes three components:
the first component uses vanilla 3D convolutional layers to extract lung image features,
the second component consists of two 3D residual blocks, which perform element conversion on the 3D feature maps,
and the third component gradually extracts the information in the 3D feature map through 3D max-pooling, and outputs the probability of COVID-19.
In \cite{song2020deep}, Song \emph{et al}. collected 1990 CT images, including 777 images from 88 patients with COVID-19, 505 images from 100 patients with bacterial pneumonia, and 708 images from 86 healthy people.
They proposed a DRE-Net DL model based on a pre-trained ResNet50 structure and a functional pyramid network.
The DRE-Net model extracts the top-$K$ lesion features from each CT image to predict the classification of patients with COVID-19.

The lack of large-scale data sets is the main challenge that hinders the implementation of AI-based CT image inspection and affects diagnostic performance.
To address these challenges, strategies such as transfer learning, data augmentation, and ``Human-In-The-Loop'' were used in \cite{zhao2020covid, shan2020lung}.
In \cite{zhao2020covid}, Zhao \emph{et al}. provided a public COVID-19 CT scan data set, including 275 COVID-19 cases and 195 non-COVID-19 cases.
They used data augmentation and TL methods to alleviate the shortage of training data.
In terms of data augmentation, they used transformation operations to expand the training data set, such as random transformation, cropping, and rotation.
In terms of TL, they pre-trained the DenseNet model \cite{huang2017densely} on the chest X-ray data set \cite{wang2017hospital}, and then used the pre-trained model to predict COVID-19.
In addition, the strategy of ``Human-In-The-Loop'' was adopted to reduce the workload of radiologists when annotating training samples \cite{shan2020lung}.
The radiologists annotated a small portion of training samples in the first batch of training.
Then, they manually corrected the segmentation results in the second batch and used them as annotations for the images.
Iterative training is performed in this way to complete the annotation of all training samples.

It is commendable that several works provide open-source code of the designed models and online COVID-19 CT image inspection systems.
For example, Li \cite{li2020artificial}, Zheng \cite{zheng2020deep}, and Zhao \cite{zhao2020covid} published the proposed DL models on GitHub \cite{github}.
In addition, Song \emph{et al}. provided an online CT diagnosis service \cite{song2020deep}, and Wang \emph{et al}. provided a public website for uploading and testing lung CT images \cite{wang2020deep}.
In \cite{chen2020deep}, Chen \emph{et al}. developed a public online CT diagnostic system, and anyone can upload CT images for self-diagnosis.
More detailed information about AI-based CT image segmentation and classification methods is provided in Table \ref{table02} and Fig. \ref{fig04}.

\begin{table}[!ht]
\renewcommand{\arraystretch}{1.1}
\setlength{\abovecaptionskip}{4pt}
\setlength{\belowcaptionskip}{4pt}
\caption{AI-based CT image segmentation and classification methods for COVID-19 inspection.}
\centering
\small
\label{table02}
\begin{threeparttable}[b]
\tabcolsep1pt
\begin{tabular}{L{0.7in} C{0.6in} C{0.4in} C{0.6in} L{1in} C{0.65in} C{0.6in}  C{0.6in}}
\hline
Literature                             &  Data sources & Data size  & COVID-19 cases & ~~~AI methods &  ACC/AUC & Sensitivity & Specificity\\
\hline
Chen \cite{chen2020deep} \tnote{1} & private  & 35,355 &  20,886 & U-Net++&  95.24\% & 100\% & 94.0\%\\
Gozes \cite{gozes2020rapid} & private & 157 & 56 & U-Net, ResNet & 99.6\% &  98.2\% &  92.2\% \\
Huang \cite{huang2020serial} & private & 842 & 842 & U-Net & - & - & - \\
Li \cite{li2020artificial} \tnote{3}  & private & 4,356 & 1,325 & U-Net, ResNet & 96.0\% & 90.0\% & 96.0\% \\
Qi \cite{qi2020machine}     & private & 52& 52 &LR, RF &   97.0\%  & 100\% & 75.0\% \\
Shan \cite{shan2020lung}  & private & 549 & 549 & V-Net, ResNet &  - & - & - \\
Shi \cite{shi2020large}  & private & 2685 & 1658 & RF & 87.9\% & 90.7\% & 83.3\% \\
Shi \cite{shi2020deep} & private & 196 & 196 & V-Net & 89.0\% & 82.2\% & 82.8\% \\
Song \cite{song2020deep} \tnote{4}& private & 1990 & 777 & DRE-Net & 94.0\% & 93.0\% & - \\
Tang \cite{tang2020severity} & private & 176 & 176 & RF & 87.5\%  & -  &  -\\
Wang \cite{wang2020deep} \tnote{5} & private & 453 & 195 & Inception & 73.1\% &  74.0\% &  67.0\% \\
Xu \cite{xu2020deep}        & private & 618   & 219  &  ResNet & 86.7\% & - & - \\
Zhao \cite{zhao2020covid} \tnote{6} & \cite{zhao2020covid} & 470 & 275 & DenseNet & 84.7\% & 76.2\% & - \\
Zheng \cite{zheng2020deep} \tnote{7} &private & 630 & 630 & DeCoVNet & 90.1\% & 90.7\% & 91.1\% \\
\hline
\end{tabular}
 \begin{tablenotes}
     \item[1]{\scriptsize  http://121.40.75.149/znyx-ncov/index.~~~~~~~~~~} $^{2}$ {\scriptsize  https://github.com/ChenWWWeixiang/diagnosis\_covid19.}
     \item[3]{\scriptsize  https://github.com/bkong999/COVNet.git.} ~~~~~$^{4}$ {\scriptsize  http://biomed.nscc-gz.cn/server/Ncov2019.}
     \item[5]{\scriptsize  https://ai.nscc-tj.cn/thai/deploy/public/pneumonia\_ct.} ~~~~$^{6}$ {\scriptsize  https://github.com/UCSD-AI4H/COVID-CT.}
     \item[7]{\scriptsize  https://github.com/sydney0zq/covid-19-detection.}
   \end{tablenotes}
   \end{threeparttable}
\end{table}

\begin{figure}[!ht]
\setlength{\abovecaptionskip}{4pt}
\setlength{\belowcaptionskip}{4pt}
  \centering
 \includegraphics[width=5.4in]{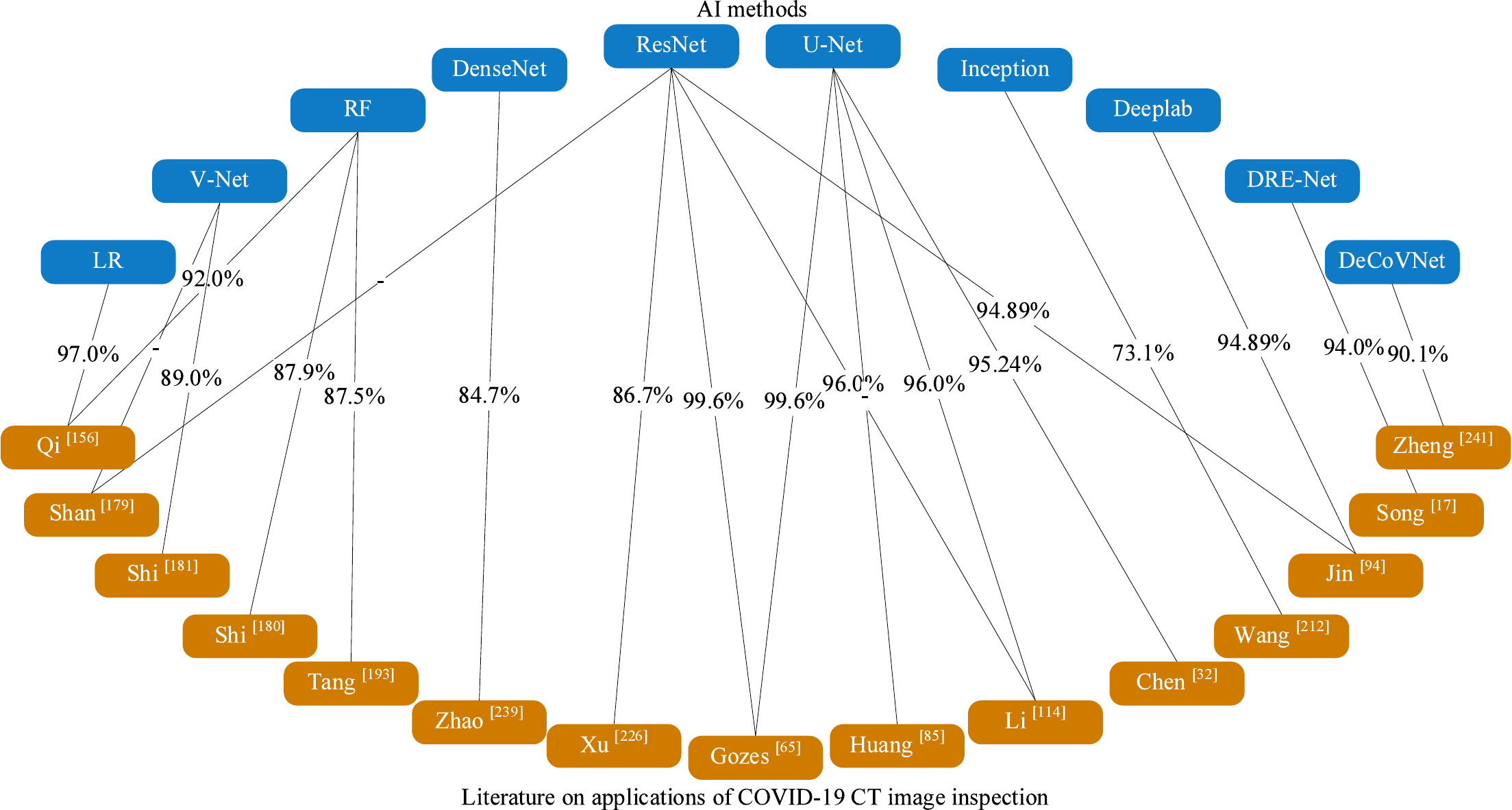}
 \caption{Relationship between AI methods and applications of COVID-19 CT image inspection.}
\label{fig04}
\end{figure}

\subsubsection{Chest X-ray Image Inspection}

\begin{figure}[!ht]
\setlength{\abovecaptionskip}{4pt}
\setlength{\belowcaptionskip}{4pt}
 \centering
 \includegraphics[width=5.45in]{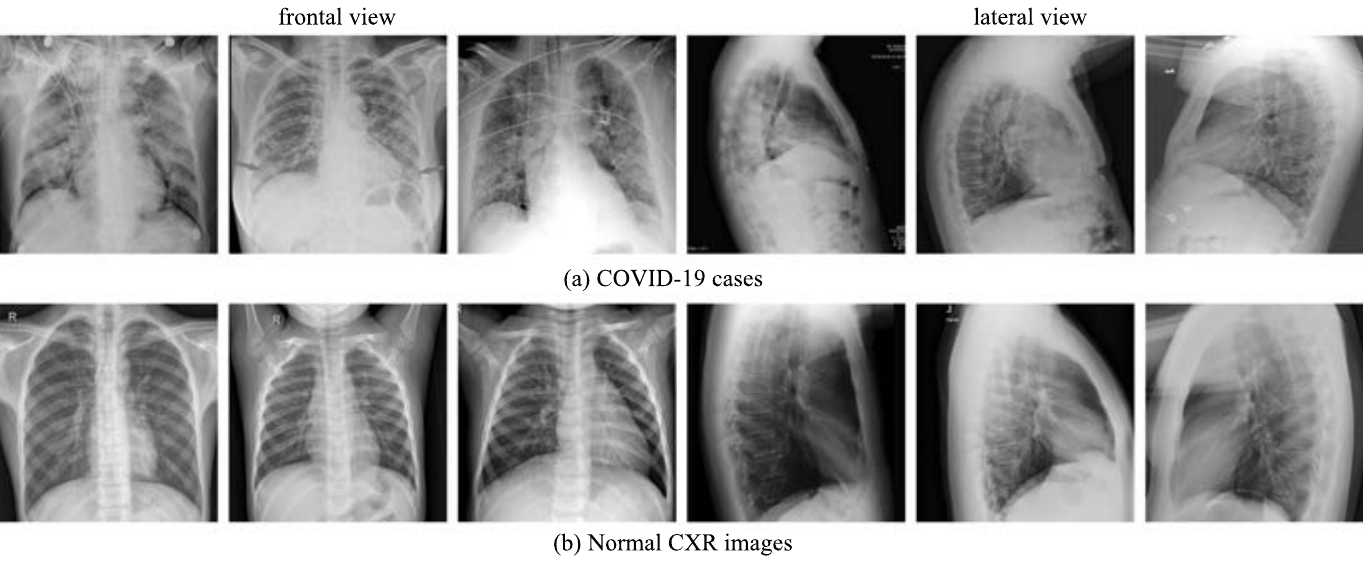}
\caption{Examples of chest X-ray images of normal and COVID-19 cases.}
\label{fig05}
\end{figure}

\begin{figure}[!ht]
\setlength{\abovecaptionskip}{4pt}
\setlength{\belowcaptionskip}{4pt}
  \centering
 \includegraphics[width=5.4in]{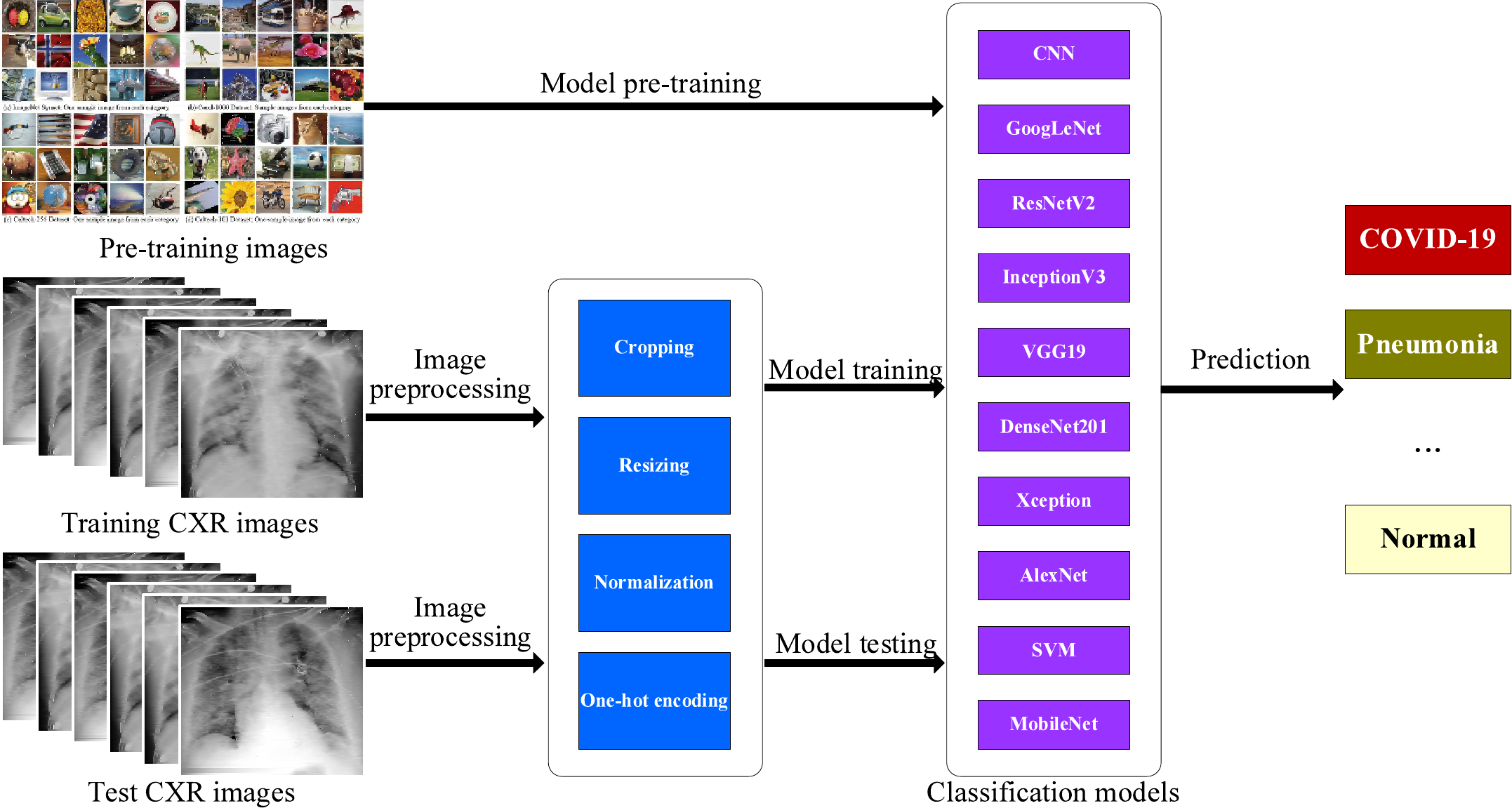}
 \caption{Representative AI architecture for CXR image classification and COVID-19 inspection.}
\label{fig06}
\end{figure}

Compared with CT images, chest X-ray (CXR) images are easier to obtain in radiological inspections.
Although CXR imaging is a typical imaging method used for the diagnosis of COVID-19, it is generally considered to be less sensitive than CT imaging.
Some CXR images of patients with early COVID-19 showed normal characteristics.
Radiological signs of COVID-19 CXR images include airspace opacity, GGO, and later mergers.
In addition, the distribution of bilateral, peripheral, and lower regions is mainly observed \cite{kaggle2020data3, mooney}.
Examples of CXR images of normal and COVID-19 cases are shown in Fig. \ref{fig05}.
The AI-based CXR image inspection usually includes steps such as data preprocessing, DL model training, and COVID-19 classification.
The representative AI architecture for CXR image classification and COVID-19 inspection is shown in Fig. \ref{fig06}.

Unlike CT images, CXR image segmentation is more challenging because the ribs are projected onto soft tissues, which is confused with image contrast.
In this way, most DL models focus on the classification of the entire CXR image, while few works are devoted to segmenting the ROIs and lung organs from CXR images.
In \cite{hassanien2020automatic}, Mahdy \emph{et al}. used a classification method to classify COVID-19 on CXR images through a multilevel threshold and SVM.
A multilevel image segmentation threshold was used to segment lung organs from the background, and then the SVM module was used to classify the infected lungs from the uninfected lungs.

Focusing on the classification of COVID-19 based on CXR images, several studies built AI-based classification models by nesting or combining existing ML and DL models.
In \cite{hemdan2020covidx}, Hemdan \emph{et al}. proposed a DL framework (called COVIDX-Net) to help radiologists automatically diagnose COVID-19 based on CXR images.
The proposed framework integrates 7 DCNN models with different structures, such as VGG19 \cite{simonyan2014very}, DenseNet201 \cite{huang2017densely}, ResNetV2 \cite{too2019comparative}, InceptionV3, InceptionResNetV2 \cite{szegedy2017inception}, Xception \cite{chollet2017xception}, and MobileNetV2 \cite{sandler2018mobilenetv2}.
Each model is separately trained on CXR images to classify the patient's status as COVID-19 positive or negative.
In \cite{sethy2020detection}, Sethy \emph{et al}. used different DCNN models in a SVM classifier to diagnose COVID-19 based on CXR images.
Eleven DCNN models are used as image feature extractors, including AlexNet, GoogLeNet, DenseNet, Inception, ResNet, VGG, XceptionNet, and InceptionResNet.
Then, the SVM classifier classifies the extracted image features to determine whether the input image is COVID-19.
In \cite{castiglioni2020artificial}, Castiglioni \emph{et al}. collected CXR data (610 images, including 324 COVID-19 cases) from Lombardy, Italy, and constructed a DCNN model to predict COVID-19.
The DCNN model is equipped with 10 CNN models, each of them uses the ResNet-50 structure and has been pre-trained on the ImageNet data set.
In \cite{apostolopoulos2020covid}, Ioannis \emph{et al}. evaluated the performance of 5 DCNN models for medical image classification, including VGG19, MobileNetV2, Inception, Xception, and InceptionResNetV2.
To address the shortcomings of the small-scale COVID-19 data set, each model is pre-trained on the ImageNet data set using the TL strategy.
In \cite{narin2020automatic}, Narin \emph{et al}. also used 3 typical DCNN models (i.e., ResNet50, InceptionV3, and InceptionResNetV2) to classify COVID-19 from a small-scale CXR data set (100 images, including 50 COVID-19 cases).

Another group of works constructed specific DL models for COVID-19 classification \cite{zhang2020covid}.
For example, Zhang \emph{et al}. proposed a new DL model, which consists of a backbone network, a classification module, and an anomaly detection module \cite{zhang2020covid}.
The backbone network extracts the features of each input CXR image.
The classification and anomaly detection modules respectively use the extracted features to generate classification scores and scalar anomaly scores.
In \cite{wang2020covid}, Wang \emph{et al}. introduced a COVID-Net DCNN model to identify COVID-19 cases based on CXR images.
The COVID-Net model uses a large number of convolutional layers in a projection-expansion-projection design pattern.
They collected 13,800 CXR images from 13,725 patients (including 183 COVID-19 patients) to establish a CXR database (called COVIDx) for training COVID-Net.
It is commendable that the authors provided an open-source code of the proposed model and the COVIDx database.

Similar to CT images, in CXR image inspection, there is also a lack of large-scale data sets for DL model training.
In \cite{maghdid2020diagnosing}, Maghdid \emph{et al}. respectively used CNN and AlexNet \cite{krizhevsky2012imagenet} models to train CXR and CT images and diagnose COVID-19 cases.
Among them, the AlexNet model is pre-trained on the ImageNet data set to perform COVID-19 classification on the data sets in \cite{cohen2020covid,BSTI, mooney}.
Unlike existing TL and image augmentation methods, Afshar \emph{et al}. designed a capsule network model (named COVID-CAPS) suitable for small-scale CXR data sets \cite{afshar2020covid}.
Each layer of the COVID-CAPS model contains multiple capsules, and each capsule represents a specific image instance at a specific position through multiple neurons.
The capsule module \cite{hinton2018matrix} uses protocol routing to capture alternative models of spatial information and attempts to reach a consensus on the existence of objects.
In this way, the protocol uses information from instances and objects to identify the relationship between them without the need for large-scale data sets.
More detailed information about AI-based methods for CXR image classification and COVID-19 inspection is shown in Table \ref{table03} and Fig. \ref{fig07}.

\begin{table}[!ht]
\renewcommand{\arraystretch}{1.1}
\setlength{\abovecaptionskip}{4pt}
\setlength{\belowcaptionskip}{4pt}
\caption{AI-based methods for CXR image classification for COVID-19 inspection.}
\centering
\small
\label{table03}
\begin{threeparttable}[b]
\tabcolsep1pt
\begin{tabular}{L{0.84in} C{0.73in} C{0.3in} C{0.6in} L{1.13in} C{0.55in} C{0.56in} C{0.56in}}
\hline
~~Literature                             &  Data sources & Data size  & COVID-19 cases & ~~~~~~~~~~~~~~AI methods &  Accuracy / AUC  & Sensitivity & Specificity\\
\hline
Afshar \cite{afshar2020covid} \tnote{1}   & \cite{cohen2020covid, kaggle-xray} & 100 & 50 & COVID-CAPS & 95.70\% & 90.00\% & 95.80\% \\
Castiglioni \cite{castiglioni2020artificial} & private & 610 & 324 & ResNet &  89.00\% & 78.00\% & 82.00\%\\
Ioannis \cite{apostolopoulos2020covid} & \cite{cohen2020covid, kaggle-xray2, kermany2018identifying} & 1427 & 224 & VGG, MobileNet, Inception, Xception, InceptionResNet  & 96.78\%  & 98.66\% & 96.46\%\\
Mahdy \cite{hassanien2020automatic} & private      & 40 & 25 & SVM & 97.48\%  & 95.28\% & 99.70\%\\
Hemdan \cite{hemdan2020covidx}    & \cite{cohen2020covid, Adrian}& 50& 25 &VGG, DenseNet, ResNet, Inception, InceptionResNet, Xception, MobileNet &   83.00\% & 91.00\% & 91.00\% \\
Maghdid \cite{maghdid2020diagnosing} &  \cite{cohen2020covid,BSTI, mooney}        & 170 & 60 & CNN, AlexNet  & 94.00\% & 100\% & 88.00\%\\
Narin \cite{narin2020automatic}  & \cite{cohen2020covid, kaggle-xray} & 100 & 50 & ResNet, Inception, InceptionResNet &   98.00\% & 96.00\% & 98.00\%\\
Sethy \cite{sethy2020detection} & \cite{cohen2020covid, kaggle-xray} & 50 & 25 & AlexNet, DenseNet, GoogLeNet, Inception, ResNet, VGG, Xception, InceptionResNet, SVM & 95.38\% & 97.29\% & 93.47\%\\
Wang \cite{wang2020covid} \tnote{2} & \cite{cohen2020covid, Chung, kaggle-xray} & 13,800 & 183 & COVID-Net & 92.6\% & 87.10\% & 96.40\% \\
Zhang \cite{zhang2020covid} &  \cite{cohen2020covid, wang2017hospital} & 1531 & 100 &  CAAD & 95.18\% & 96.00\% & 70.65\%\\
\hline
\end{tabular}
 \begin{tablenotes}
     \item[1] {\scriptsize https://github.com/ShahinSHH/COVID-CAPS.} $^2$ {\scriptsize https://github.com/lindawangg/COVID-Net.}
   \end{tablenotes}
   \end{threeparttable}
\end{table}

\begin{figure}[!ht]
\setlength{\abovecaptionskip}{4pt}
\setlength{\belowcaptionskip}{4pt}
  \centering
 \includegraphics[width=3.4in]{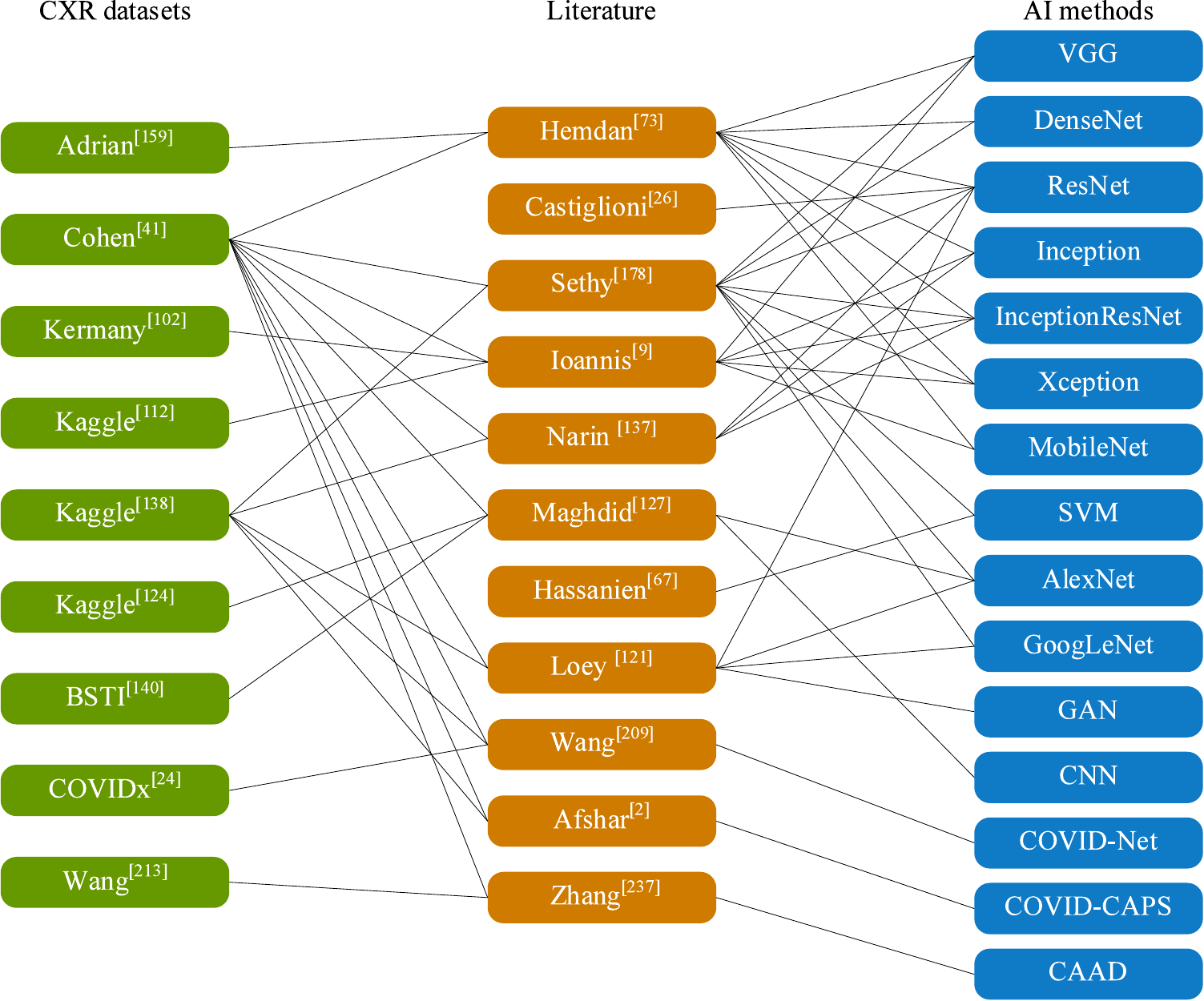}
 \caption{Relationship among CXR data sets, AI methods, and applications of CXR image inspection.}
\label{fig07}
\end{figure}

\subsection{Other Detection Technologies}

In addition to RT-PCR detection and image inspection techniques, some noninvasive measurement methods have also been used for the diagnosis of COVID-19, including cough sound judgment and breathing pattern detection.

(1) Monitoring COVID-19 through AI-based cough sound analysis.

In \cite{schuller2020covid}, Schuller \emph{et al}. discussed the potential application of computer audition (CA) and AI in analyzing the cough sounds of patients with COVID-19.
They first analyzed the ability of CA to automatically recognize and monitor speech and cough under different semantics (such as breathing, dry and wet coughing or sneezing, speech during colds, eating behaviors, drowsiness, or pain).
Then, they applied the CA technology to the diagnosis and treatment of patients with COVID-19.
However, due to the lack of available data sets and annotation information, there are few studies on this technology in diagnosing COVID-19.
It is encouraging that Orlandic \emph{et al}. provided a COUGHVID cough dataset for COVID-19 diagnosis, which contains 20,000 crowdsourced cough records \cite{orlandic2020coughvid}.
The cough records are labeled by experienced pulmonologists, and medical abnormalities are accurately diagnosed, thus providing a high-quality training dataset for MI/AI methods.
Similarly, in \cite{iqbal2020active}, Iqbal \emph{et al}. also discussed an abstract framework that uses the voice recognition function of mobile applications to capture and analyze the cough sounds of suspicious persons to determine whether they are healthy or suffering from a respiratory disease.
In \cite{wang2020abnormal}, Wang \emph{et al}. analyzed the respiratory patterns of patients with COVID-19 and other breathing patterns of patients with influenza and common cold.
In addition, they proposed a respiratory simulation model (called BI-AT-GRU) for COVID-19 diagnosis.
The BI-AT-GRU model includes a GRU neural network with a bidirectional and attention mechanism, and can classify 6 types of clinical respiratory patterns, such as Eupnea, Tachypnea, Bradypnea, Biots, Cheyne-Stokes, and Central-Apnea.
To facilitate the analysis of COVID-19 cough sounds, we further collect 8 groups of human cough sound datasets containing patients with confirmed COVID-19, as shown in Table \ref{table10}.

(2) COVID-19 diagnosis based on non-invasive measurements.

In \cite{maghdid2020novel}, Maghdid \emph{et al}. designed an abstract framework for COVID-19 diagnosis based on smart phone sensors.
In the proposed framework, smart phones can be used to collect the disease characteristics of potential patients.
For example, the sensor can acquire the patient's voice through the recording function and the patient's body temperature through the fingerprint recognition function.
Then, the collected data are submitted to an AI-supported cloud server for disease diagnosis and analysis.

\section{Virology and Pathogenesis}
\label{section3}
The virology and pathogenesis of SARS-CoV-2 is the most important scientific research in the fields of biology and medicine.
Scientists have analyzed the virus characteristics of SARS-CoV-2 through proteomics and genomic studies \cite{andersen2020proximal, hoffmann2020sars, lu2020genomic}.
In the field of virology, the origin and classification of SARS-CoV-2, physical and chemical properties, receptor interactions, cell entry, and the ecology and genomic variation of SARS-CoV-2 have been studied \cite{zhou2020pneumonia, lu2020genomic, wolfel2020virological}.
We mainly discuss the contribution of AI in the pathological research of SARS-CoV-2 from the perspective of proteomics and genomics.

\subsection{Proteomics}
\begin{figure}[!ht]
\setlength{\abovecaptionskip}{4pt}
\setlength{\belowcaptionskip}{4pt}
  \centering
 \includegraphics[width=5.4in]{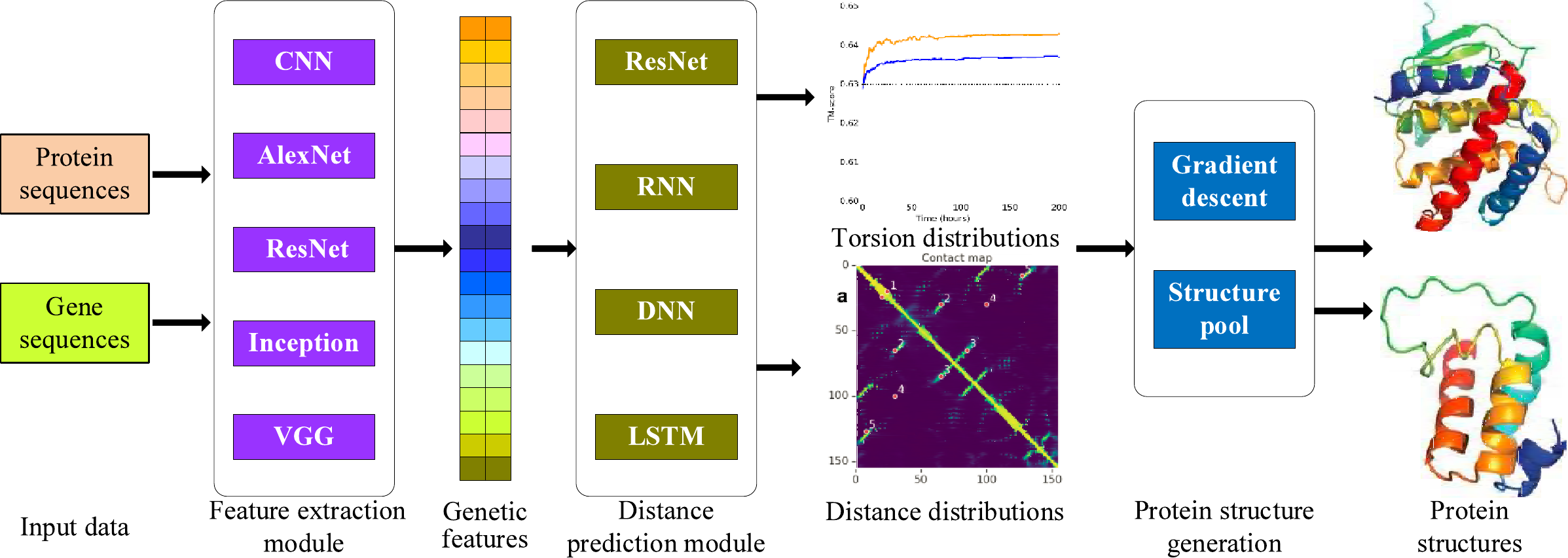}
 \caption{Representative AI architecture used for protein structure predication.}
\label{fig08}
\end{figure}

Since the advent of SARS-CoV-2, there have been a large number of research achievements in proteomics.
Five types of structural proteins of SARS-CoV-2 have been confirmed, including nucleocapsid (N) proteins, envelope (E) proteins, membrane (M) proteins, and spike (S) proteins \cite{zhavoronkov2020potential, ortega2020role, walls2019unexpected}.
Other proteins translated in the host cells essential for virus replication, such as non-structural protein 5 (NSP5) and 3C-like protease (3CLpro), have also attracted the attention of researchers.
In addition, several studies have shown that SARS-CoV-2 uses the human Angiotensin-Converting Enzyme 2 (ACE2) to enter the host \cite{hoffmann2020sars, zhou2020pneumonia}.
In this field, AI techniques are used to predict protein structures and analyze the interaction network between proteins and drugs.
The representative AI architecture used for protein structure prediction is shown in Fig. \ref{fig08}.

In \cite{senior2020improved, senior2019protein}, Senior \emph{et al}. used DL models to implement the AlphaFold system for protein structure prediction.
The AlphaFold system uses the ResNet model \cite{he2016deep} to analyze the covariance and amino acid residue contacts in the homologous gene sequences, and predict the corresponding protein structures.
The AlphaFold system consists of a feature extraction module and a distance prediction neural network.
The feature extraction module is responsible for searching for protein sequences similar to the input protein sequences and constructing multiple sequence alignments (MSA).
The module simultaneously generates residual position and sequence contour features, and inputs the output of 485 feature parameters into the distance prediction neural network.
The distance prediction neural network is a two-dimensional (2D) ResNet structure, which is responsible for accurately predicting the distance between all residue pairs in every two protein sequences.
The authors added a one-dimensional output layer to the network to predict the accessible surface area, distance map, and secondary structure of each residue.
Finally, the generated potential is optimized by gradient descent to generate protein structures.

Based on \cite{senior2020improved, senior2019protein}, Jumper \emph{et al}. used the AlphaFold system to predict the structure of SARS-CoV-2 membrane proteins \cite{deepmind}.
They published the predicted protein structures such as 3a, Nsp2, Nsp4, Nsp6, and papain-like proteases.
Although the structure of these proteins has not been verified by clinical trials, this publication allows researchers to quickly conduct SARA-CoV-2 studies.
In \cite{ortega2020role}, Ortega \emph{et al}. used a computational method to detect changes in the S1 subunit of the spike receptor-binding domain and identify mutations in the SARS-CoV-2 spike protein sequence, which may be beneficial for studying human-to-human transmission.
They collected sequences from the Protein Data Bank (PDB) \cite{berman2003announcing} and used the SWISS-MODEL software \cite{arnold2006swiss} to construct a SARS-CoV-2 spike protein model.
Then, the Z-dock software \cite{pierce2014zdock} was used to dock the spike protein and ACE2, and a clustering algorithm was used to cluster the docking results.
The work indicated that the SARS-CoV-2 spike protein has a higher affinity for human ACE2 receptors.

Another branch of AI-assisted proteomics research involves finding new compounds and drug candidates for the treatment of COVID-19 by building an interactive network and knowledge graph between proteins and drugs.
Please see Section 4 for details.

\subsection{Genomics}
Genomics is mainly used in SARS-CoV-2 to analyze the origin of SARS-CoV-2, vaccine development, and PT-PCR detection.
Various AI algorithms are applied to compare the similarity of gene sequences, gene fragments, and miRNA predictions \cite{randhawa2020machine, demirci2020computational}.
In \cite{randhawa2020machine}, Randhawa \emph{et al}. used different ML methods to analyze the pathogen sequence of COVID-19, and identified the inherent characteristics of the viral genomes, thereby rapidly classifying new pathogens.
They collected the complete reference genome of the COVID virus from NCBI \cite{ncbi2020data}, the bat $\beta$-coronavirus from GISAID \cite{gisaid2020data}, and all available virus sequences from Virus-Host DB \cite{mihara2016linking}.
By using chaotic game representation, each genomic sequence was mapped to the corresponding genomic signal in a discrete digital sequence \cite{jeffrey1990chaos}.
In addition, the amplitude spectrum of these genomic signals was calculated by using a discrete Fourier transform.
On this basis, they used 6 ML classification models to train the above sequence distance matrix and compared their performance.
Finally, they conducted the trained ML models on 29 COVID-19 sequences to classify COVID-19 pathogens.
The results of this work support the hypothesis that COVID-19 originated in bats and its classification as a $\beta$-coronavirus.

In addition, the amplitude spectrum of these genomic signals is calculated by using a discrete Fourier transform.
On this basis, they used 6 ML classification models (such as linear discriminant, linear SVM, quadratic SVM, fine KNN, subspace discriminant, and subspace KNN) to train the sequence distance matrix and compared their performance.
Finally, they performed a well-trained ML model on 29 COVID-19 sequences to classify COVID-19 pathogens.
The results of this work support the following hypothesis: COVID-19 originated in bats and is classified as type $\beta$-coronavirus.

In \cite{demirci2020computational}, based on three ML methods (such as DT, Naive Bayes, and RF), Demirci \emph{et al}. performed miRNA prediction on the SARS-CoV-2 genome, and identified miRNA-like hairpins and microRNA-mediated SARS-CoV-2 infection interactions.
They collected the complete COVID-19 genome from NCBI \cite{ncbi2020data} and human-mature miRNA sequences from miRBase \cite{kozomara2019mirbase}.
The genomic sequences are transcribed and divided into multiple overlapping fragments, which are folded into a secondary structure to extract the hairpin structure.
On this basis, the authors used the three ML methods to predict the category of each hairpin and determine the similarity between the hairpins and human miRNA.
They searched for mature miRNA targets in human and SARS-CoV-2 genes, and analyzed the potential interactions between SARS-CoV-2 miRNAs and human genes and between human miRNAs and SARS-CoV-2 genes.
Finally, the gene ontology of SARS-CoV-2 miRNA targets in human genes was analyzed, and the PANTHER classification system was used to evaluate the similarity between SARS-CoV-2 miRNA candidates and the mature miRNAs of any known organism \cite{mi2012panther}.

In \cite{metsky2020crispr}, Metsky \emph{et al}. used genomic and AI technologies to design nucleic acid detection assays and improve the current SARS-CoV-2 RT-PCR test.
They developed a CRISPR tool that uses enzymes to edit the genome by cutting specific genetic code chains, and used different ML methods to predict the diversity of the target genome.
The authors designed an RT-PCR test method through the CRISPR tool, which can effectively detect 67 respiratory viruses, including SARS-CoV-2.

\section{Drug and Vaccine Development}
\label{section4}
Based on proteomics and genomics research, a variety of drug and vaccine development programs have been proposed for SARS-CoV-2 and COVID-19.
The application of AI in the development of new drugs and vaccines is one of the main contributions in smart medicine, and it plays an important role in the battle against COVID-19.

\subsection{Drug Development}
\begin{figure}[!ht]
\setlength{\abovecaptionskip}{4pt}
\setlength{\belowcaptionskip}{4pt}
  \centering
 \includegraphics[width=5.4in]{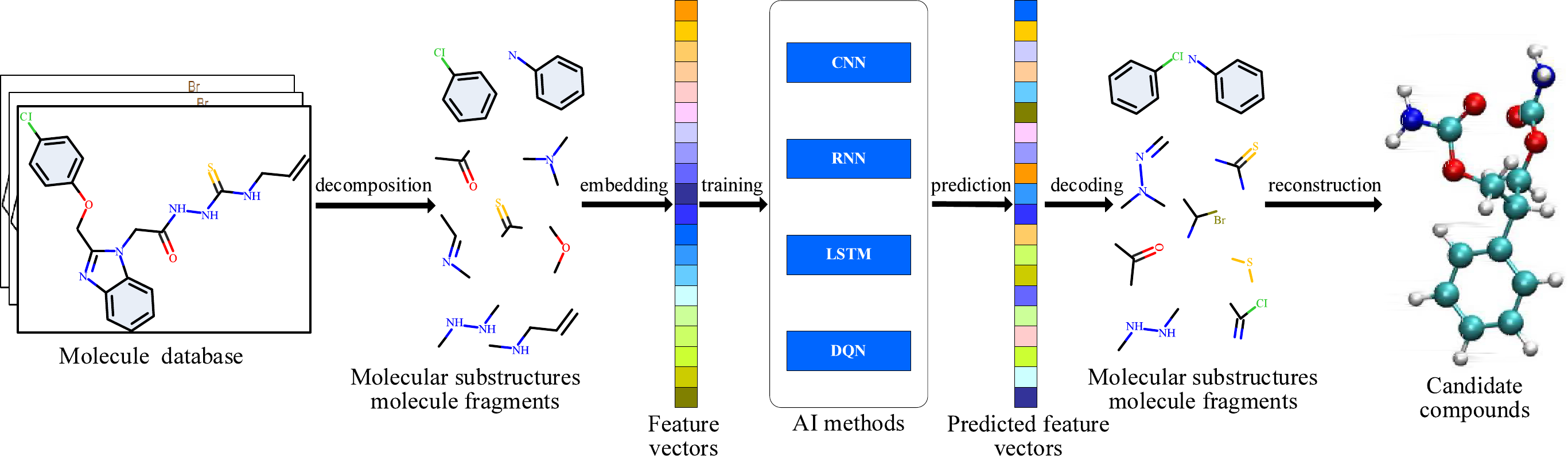}
 \caption{Representative AI architecture used for new drug-like compound discover.}
\label{fig09}
\end{figure}

In the field of drug development, AI technologies can screen existing drug candidates for COVID-19 by analyzing the interaction between existing drugs and COVID-19 protein targets.
In addition, AI technologies can also help discover new drug-like compounds against COVID-19 by constructing new molecular structures that inhibit proteases at the molecular level.
The representative AI architecture used for new drug-like compound discover is shown in Fig. \ref{fig09}.

Drug development can be divided into small-molecule drug discovery and biological product development.
The discovery of small-molecule drugs mainly focused on chemically synthesized small molecule active substances, which can be made into small-molecule drugs through chemical reactions between different organic and inorganic compounds.
One group of AI-based drug development focuses on discovering new drug-like compounds at the molecular level.
In \cite{beck2020predicting, shin2019self}, Beck \emph{et al}. proposed a DL-based molecule transformer-drug target interaction (MT-DTI) model to predict potential drug candidates for COVID-19.
The MT-DTI model uses the SMILES strings and amino-acid sequences to predict the target proteins with a 3D crystal structure.
The authors collected the amino-acid sequences of 3C-like proteases and related antiviral drugs and drug targets from NCBI \cite{ncbi2020data}, Drug Target Common (DTC) \cite{tanoli2018drug}, and BindingDB \cite{liu2007bindingdb} databases.
In addition, they used a molecular docking and virtual screening tool (AutoDock Vina \cite{trott2010autodock}) to predict the binding affinity between 3,410 drugs and SARS-CoV-2 3CLpro.
The experimental results provided 6 potential drugs, such as Remdesivir, Atazanavir, Efavirenz, Ritonavir,  Dolutegravir, and Kaletra (lopinavir/ritonavir).
Note that Remdesivir shows promising in clinical trials.
In \cite{moskal2020suggestions}, Moskal \emph{et al}. used AI methods to analyze the molecular similarity between anti-COVID-19 drugs (termed ``parents'') and drugs involving similar indications to screen out second-generation drugs (termed ``progeny'') of COVID-19.
They first used the Mol2Vec \cite{jaeger2018mol2vec} method to convert the molecular structure of the parent drugs into a high-dimensional vector space, treated the drug molecule as a  ``sentence'', and mapped its molecular substructure to a ``word''.
Then, they used the VAE \cite{skalic2019shape} model to generate SMILES strings with 3D shape and pharmacodynamic properties to the given seed molecule \cite{gomez2018automatic}.
In addition, CNN, LSTM, and MLP models are used to generate the corresponding SMILES strings and molecules.
The authors selected 71 parent drugs from the literature as seed molecules, and 4456 drugs from ZINC \cite{zinc2020data} and ChEMBL as candidate progeny drugs \cite{gaulton2012chembl}.

In \cite{bung2020novo}, Bung \emph{et al}. committed to the development of new chemical entities of SARS-CoV-2 3CLpro based on DL technology.
They constructed an RL-based RNN model to classify protease inhibitor molecules and obtained a smaller subset that favored the chemical space.
Then, they collected 2515 protease inhibitor molecules in the SMILES format from the ChEMBL database as training data, where each SMILES string is regarded as a time series, and each position or symbol is regarded as a time point.
The output of small molecules is docked to the 3CLpro structure with minimal energy and ranked according to the virtual screening score obtained by selecting anti-SARS-CoV-2 candidates \cite{trott2010autodock}.
In \cite{tang2020ai}, Tang \emph{et al}. analyzed 3CLpro with a 3D structure similar to SARS-CoV, and evaluated it as an attractive target for the development of anti-COVID-19 drugs.
They proposed an advanced deep Q learning network (called ADQN-FBDD) to generate potential lead compounds of SARS-CoV-2 3CLpro.
They collected 284 molecules reported as SARS-CoV-2 3CLpro inhibitors.
An improved BRICS algorithm \cite{degen2008art} was used to split these molecules to obtain the target fragment library of SARS-CoV-2 3CLpro.
Then, the proposed ADQN-FBDD model trains each target fragment and predicts the corresponding molecules and lead compounds.
Through the proposed Structure-Based Optimization Policy (SBOP), they finally obtained 47 derivatives that inhibit SARS-CoV-2 3CLpro from these lead compounds, which are regarded as potential anti-SARS-CoV-2 drugs.

Another group of studies focuses on screening candidate biological products of COVID-19.
Biological products are protein products with therapeutic effects, which mainly bind to specific cell receptors involved in the disease process.
Biological products are prepared from microbial cells (such as genetically modified bacteria, yeast, or mammalian cell strains) through a biotechnology process.
In \cite{hu2020prediction}, Hu \emph{et al}. established a multi-task DL model to predict the possible binding between potential drugs and SARS-CoV-2 protein targets, so as to select drugs that can be used for SARS-CoV-2.
They first collected 8 SARS-CoV-2 viral proteins from GHDDI \cite{ghddi2020data} as potential targets.
The proposed DL model is based on the AtomNet model \cite{wallach2015atomnet, stepniewska2018development} and includes a shared layer to learn the joint representation of all tasks and a task processing layer to perform specific tasks.
By fine-tuning the DL model using a coronavirus-specific data set, the model can predict the possible binding between the drugs and the protein targets and output the binding affinity score.
Based on existing studies, RdRp, 3CLpro, and papain-like protease have been confirmed as the three principal targets of SARS-CoV-2 \cite{walls2019unexpected, ou2020characterization, gilman2019structure}.
Based on the prediction results \cite{li2020therapeutic, wang2020remdesivir}, the authors selected the top 10 potential drugs with a high likelihood of inhibiting each target.
In \cite{kadioglu2020identification}, Kadioglu \emph{et al}. used High-Performance Computing (HPC), virtual drug screening, molecular docking, and ML technologies to identify SARS-CoV-2 drug candidates.
After performing virtual drug screening and molecular docking, two supervised ML models(such as NN and Naivebayes) are used to analyze clinical drugs and test compounds to construct corresponding drug likelihood prediction models.
Several approved drugs were selected as SARS-CoV-2 drug candidates, including drugs for hepatitis C virus (HCV), enveloped ssRNA virus, and other infectious diseases.

Facing the known COVID-19 protease target 3CLpro, Zhavoronkov \emph{et al}. designed a small-molecule drug-discovery pipeline to produce 3CLpro inhibitors, and used 3CLpro's crystal structure, homology modelling and co-crystallized fragments to generate 3CLpro molecules \cite{zhavoronkov2020potential}.
They collected the crystal structure of COVID-19 3CLpro from \cite{yang2005design} and constructed a homology model.
At the same time, molecules activity to various proteases were extracted from \cite{cai2020data, gaulton2012chembl} and constituted a protease peptidomimetic data set with 5,891 compounds.
Then, they used 28 ML methods (such as GAE, GAN, and GA) and RL strategies to separately train the input data sets (e.g., crystal structure, homology model, and co-crystal ligands), and generated new molecular structures with a high score.
In \cite{hofmarcher2020large}, Hofmarcher \emph{et al}. used the ChemAI DL model \cite{preuer2019interpretable} based on the SmilesLSTM structure \cite{hochreiter1997long} to test the resistance of the molecules to COVID-19 protease.
They collected 3.6 million molecules from ChEMBL \cite{gaulton2012chembl}, ZINC \cite{zinc2020data}, and PubChem \cite{kim2016pubchem} databases and formed a training data set.
Then, the ChemAI model is trained on the data set in a multi-task parallel training way, where the output neurons of the model represent the biological effects of the input molecules.
The authors used the ChemAI model to predict the inhibitory effects of these molecules on the 3CLpro and PLpro proteases of COVID-19.
These molecules have a binding, inhibitory, and toxic effect on the targets.
A list of COVID-19 drug development methods based on AI technology is provided in Table \ref{table04}.

\begin{table}[!ht]
\renewcommand{\arraystretch}{1.1}
\setlength{\abovecaptionskip}{4pt}
\setlength{\belowcaptionskip}{4pt}
\caption{AI-based COVID-19 drug development methods.}
\centering
\small
\label{table04}
\begin{threeparttable}[b]
\tabcolsep1pt
\begin{tabular}{L{1.0in} L{0.8in} L{1.5in} C{1.0in} C{0.8in}}
\hline
Literature           &  AI methods & ~~~Role of AI methods & COVID-19 targets & Potential drugs\\
\hline
Tang \cite{tang2020ai} \tnote{1} & RL, DQN & predict molecules and lead compounds for each target fragment &  3CLpro & 47 compounds\\
Zhavoronkov \cite{zhavoronkov2020potential} \tnote{2}  & 28 ML & generate new molecular structures for 3CLpro  &3CLpro & 100 molecules \\
Bung \cite{bung2020novo} & RNN, RL  & classify protease inhibitor molecules  & 3CLpro & 31 compounds\\
Hofmarcher \cite{hofmarcher2020large} \tnote{3}  &ChemAI & predict inhibitory effects of molecules on COVID-19 proteases & 3CLpro, PLpro  & 30,000 molecules\\
Kadioglu \cite{kadioglu2020identification}  & NN, Naive bayes & construct drug likelihood prediction model & spike protein, $\cdots$ & 3 drugs \\
Hu \cite{hu2020prediction}  & DL & predict binding between drugs and protein targets &  3CLpro, RdRp, $\cdots$ & 10 drugs \\
Beck \cite{beck2020predicting, shin2019self} & MT-DTI & predict binding affinity between drugs and protein targets & 3CLpro, RdRp, helicase, $\cdots$ & 6 drugs \\
Moskal \cite{moskal2020suggestions}  & VAE, CNN, LSTM, MLP & generate SMILES strings and molecules & - & 110 drugs \\
\hline
\end{tabular}
 \begin{tablenotes}
     \item[1] {\scriptsize https://github.com/tbwxmu/2019-nCov.} $^{2}$ {\scriptsize https://www.insilico.com/ncov-sprint.}
     \item[3] {\scriptsize https://github.com/ml-jku/sars-cov-inhibitors-chemai.}
   \end{tablenotes}
\end{threeparttable}
\end{table}

\subsection{Vaccine Development}
Currently, there are 3 types of COVID-19 vaccine candidates, such as whole virus vaccines, recombinant protein subunit vaccines, and nucleic acid vaccines \cite{chen2020sars, zhang2020progress}.
AI technology has participated in the design and development of COVID-19 vaccine.
Compared with the explicit applications in other fields, AI technology is usually used in the sub-processes of vaccine development in an implicit manner.

The AI algorithms of netMHC and netMHCpan are used to develop COVID-19 vaccines for epitope prediction \cite{jurtz2017netmhcpan, herst2020effective, Ward2020an}.
In \cite{herst2020effective}, Herst \emph{et al}. obtained the SARS-CoV-2 protein sequences from GenBank and used the MSA algorithm to trim the nucleocapsid phosphoprotein sequences into possible peptide sequences.
On this basis, they used netMHC and netMHCpan AI algorithms to train and predict peptide sequences \cite{andreatta2016gapped, jurtz2017netmhcpan}.
The pan variant of netMHC integrates the in-vitro objects of 215 HLAs for prediction.
Finally, they used the average value of the ANN, SVM, netMHC and netMHCpan methods to calculate vaccine candidates.
In \cite{Ward2020an}, Ward \emph{et al}. downloaded the SARS-CoV-2 nucleotide sequences from NCBI \cite{ncbi2020data} and GISAID \cite{gisaid2020data} databases, and generated a consensus sequence for each SARS-CoV-2 protein.
The sequences can be used as references for prediction, specificity, and epitope mapping analysis.
Next, the authors used different epitope prediction tools to predict B cell epitopes and map them to the amino acid sequences of each gene.
On this basis, they used the AI-based netMHCpan algorithm to predict HLA-1 peptides and obtained a total of 2,915 alleles in all peptide lengths.
The BLASTp tool \cite{altschul1990basic} was used to map the short amino acid epitope sequences to the canonical sequences of SARS-CoV-2 proteins.
Finally, the authors provided an online tool that provides functions of SARS-CoV-2 genetic variation analysis, epitope prediction, coronavirus homology analysis, and candidate proteome analysis.

In \cite{ong2020covid}, Ong \emph{et al}. used ML and Reverse Vaccinology (RV) methods to predict and evaluate potential vaccines for COVID-19.
They used RV to analyze the bioinformatics of pathogen genomes to identify promising vaccine candidates.
They obtained the SARS-CoV-2 sequences and all the proteins of six known human coronavirus strains from the NCBI \cite{ncbi2020data} and UniProt \cite{bairoch2005universal} databases.
Then, they used Vaxign and Vaxign-ML \cite{he2010vaxign, ong2020vaxign} to analyze the complete proteome of the coronavirus and predicted its service biological characteristics.
Next, they used LR, SVM, KNN, RF, and XGBoost methods to improve the Vaxign-ML model, and predicted the protein levels of all SARS-CoV-2 proteins.
The nsp3 protein was selected for phylogenetic analysis, and the immunogenicity of nsp3 was evaluated by predicting T cell MHC-I and MHC-II and linear B-cell epitopes.

In \cite{qiao2020personalized}, Qiao \emph{et al}. used DL to predict the patient's neoantigen mutation and identified the best T-cell epitope for the peptide-based COVID-19 vaccines.
They first sequenced the diseased cells in the patient's blood and extracted 6 human leukocyte antigen (HLA) types and T-cell receptor (TCR) sequences.
Then, they proposed the DeepNovo model to train patient's immune peptides and to identify the best T-cell epitope set based on a person's HLA alleles and immune peptide group information.
The DeepNovo model uses LSTM and RNN structures to capture sequence patterns in peptides or proteins, and predicts HLA peptides from conserved regions of the virus, thereby predicting new mutant antigens in patients.
In addition, they used the IEDB tool \cite{vita2010immune} to predict the immunogenicity of 177 peptides.
They suggested designing an epitope-based COVID-19 vaccine specifically for each person based on his/her HLA alleles.

The prediction of immune stimulation ability is an important part of vaccine design \cite{rapin2010computational, rahman2020epitope}.
Different ML methods and position-specific scoring matrices (PSSM) are usually used to predict epitope and immune interactions, thereby predicting the production of adaptive immunity in the target host.
In \cite{rahman2020epitope}, Rahman \emph{et al}. used immuno-informatics and comparative genomic methods to design a multi-epitope peptide vaccine against SARS-CoV-2, which combines the epitopes of S, M, and E proteins.
They used the Ellipro antibody epitope prediction tool \cite{ellipro} to predict linear B-cell epitopes on the S protein.
In addition, Sarkar \emph{et al}. studied the epitope-based vaccine design against COVID-19 and used the SVM method to predict the toxicity of the selected epitopes \cite{sarkar2020essential}.
In \cite{prachar2020covid}, Prachar \emph{et al}. used 19 epitope-HLA combination prediction tools including IEDB, ANN, and PSSM algorithms to predict and verify 174 SARS-CoV-2 epitopes.

\section{Epidemic and Transmission Prediction}
\label{section5}
Thanks to the developed information and multimedia technology, the outbreak and spread of COVID-19 were reported timely and accurately.
The number of suspected, confirmed, cured, and dead COVID-19 cases in each country/region is announced in real-time.
In addition, passenger travel trajectories and related big data are shared for scientific research.
Based on a wealth of data, numerous researchers have participated in the prediction, spread, and tracking of the COVID-19 outbreak.

\subsection{Patient Mortality and Survival Rate Prediction}
\begin{figure}[!ht]
\setlength{\abovecaptionskip}{4pt}
\setlength{\belowcaptionskip}{4pt}
  \centering
 \includegraphics[width=5.4in]{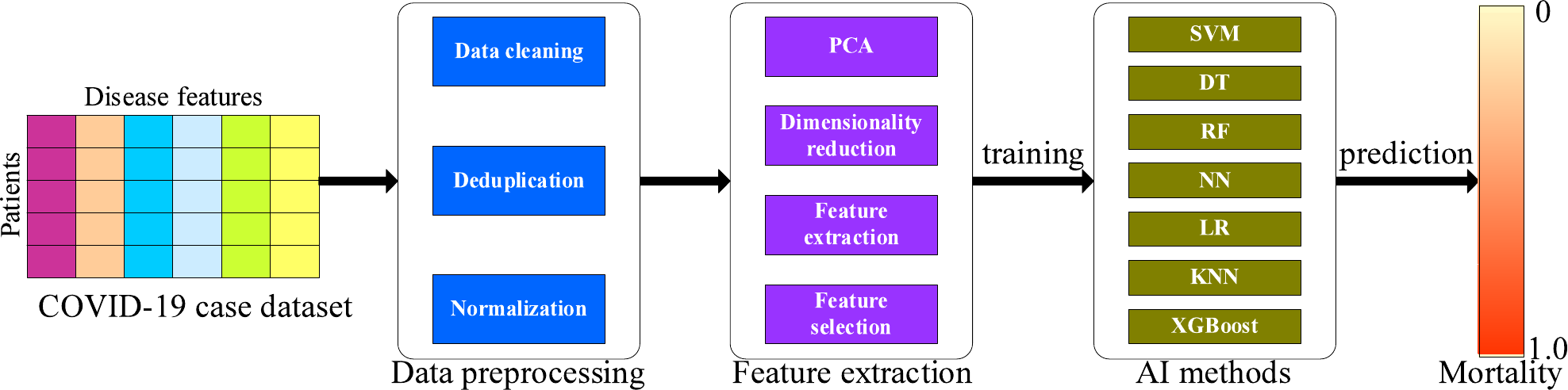}
 \caption{Representative AI architecture used for patient mortality and survival rate prediction}.
\label{fig10}
\end{figure}

Researchers collected clinical COVID-19 case data and used different AI methods to extract important features and to predict the mortality and survival rate of patients with COVID-19.
The representative AI architecture used for patient mortality and survival rate prediction is shown in Fig. \ref{fig10}.

In \cite{pourhomayoun2020predicting}, Pourhomayoun \emph{et al}. used 6 AI methods to predict the mortality of patients with COVID-19.
They used public data of patients with COVID-19 from 76 countries/regions around the world \cite{xu2020epidemiological}, and counted 112 features, including 80 medical annotations and disease features and 32 features from the patients' demographic and physiological data.
Based on the filtering method and wrapper method, 42 best features were extracted, such as demographic features, general medical information, and patient symptoms.
On this basis, 6 AI methods (such as SVM, NN, RF, DT, LR, and KNN) are used to predict the mortality of patients with COVID-19.
In \cite{Sarkar2020a}, Sarkar \emph{et al}. used the RF model to analyze the records of 433 patients with COVID-19 from Kaggle \cite{Su2020data}, and identified the important features and their impact on mortality.
Experimental results show that patients over 62 years of age have a higher risk of death.
In \cite{yan2020machine, yan2020prediction}, Yan \emph{et al}. analyzed a blood sample data set of 404 patients with COVID-19 in Wuhan, China, and used the XGBoost classification method \cite{chen2016xgboost} to select three important biomarkers to predict the survival rate of individual patients.
Experimental results with an accuracy of 90\% indicate that higher LDH levels seem to play an important role in distinguishing the most critical COVID-19 cases.

\subsection{Outbreak and Transmission Prediction}
BlueDot \cite{bluedot} and Metabiota \cite{metabiota} are two AI companies that made accurate predictions for the COVID-19 outbreak.
BlueDot collected large-scale heterogeneous data from various sources, such as news reports, global ticketing data, animal diseases, global infectious disease alerts, and real-time climate conditions.
Then, it used filtering tools to narrow its focus;
used various ML and Natural Language Processing (NLP) techniques to detect, mark, and display the potential risk frequency of COVID-19;
and finally predicted the outbreak time of transmission.
It is worth mentioning that 9 days before the official announcement of the COVID-19 outbreak, BlueDot has accurately predicted the epidemic of COVID-19 and cities with a high risk of virus outbreaks.
Metabiota collected large-scale data from social and nonsocial sources (such as biology, socioeconomic, political, and environmental data), and used technologies such as AI, ML, big data, and NLP to accurately predict the outbreak, spread, and intervention measures of COVID-19.
More AI-based methods used for COVID-19 outbreak and transmission prediction are shown in Table \ref{table06}.

\begin{table}[!ht]
\renewcommand{\arraystretch}{1.1}
\setlength{\abovecaptionskip}{4pt}
\setlength{\belowcaptionskip}{4pt}
\caption{AI methods used for COVID-19 outbreak and transmission prediction.}
\centering
\small
\label{table06}
\tabcolsep1pt
\begin{tabular}{L{1.05in} L{1.75in} L{1.5in} L{1.1in} }
\hline
Literature        & Data sources         &  Methods & Country/region   \\
\hline
Huang \cite{huang2020multiple} & Yang \cite{yang2020clinical}, WHO \cite{WHO2020data}     & CNN, LSTM, MLP, GRU & China  \\
Hu \cite{hu2020artificial, hu2020evaluating} &The Paper \cite{Surging}, WHO \cite{WHO2020data} &  MAE, clustering & China \\
Yang \cite{yang2020modified} & Baidu \cite{baidu2020data} & SEIR, LSTM & China  \\
Fong \cite{fong2020finding, fong2020composite} & NHC \cite{nhc2020data} & SVM, PNN & China\\
Ai \cite{al2020optimization} & WHO \cite{WHO2020data, cdc2020data} & ANFIS, FPA &  China, USA  \\
Rizk \cite{rizk2020covid} & WHO \cite{WHO2020data} & ISACL-MFNN & USA, Italy, Spain \\
Giuliani \cite{giuliani2020modelling} & Italy \cite{Italy2020data} &EMTMGL & Italy \\
Marini \cite{marini2020covid, marini2020enhancing} & Swiss population & Enerpol & Switzerland\\
Lai \cite{lai2020assessing} & IATA \cite{iata2020data}, Worldpop \cite{worldpop2020data}& ML & Global\\
Punn \cite{punn2020covid} & JHU CSSE \cite{dong2020interactive} & SVR, PR, DNN, LSTM, RNN & Global\\
Lampos \cite{lampos2020tracking} &  MediaCloud \cite{mediaCloud2020data}, PHE \cite{govuk2020data}, ECDC \cite{ecdc2020data} & Transfer learning & Global \\
\hline
\end{tabular}
\end{table}

\begin{figure}[!ht]
\setlength{\abovecaptionskip}{4pt}
\setlength{\belowcaptionskip}{4pt}
  \centering
 \includegraphics[width=5.4in]{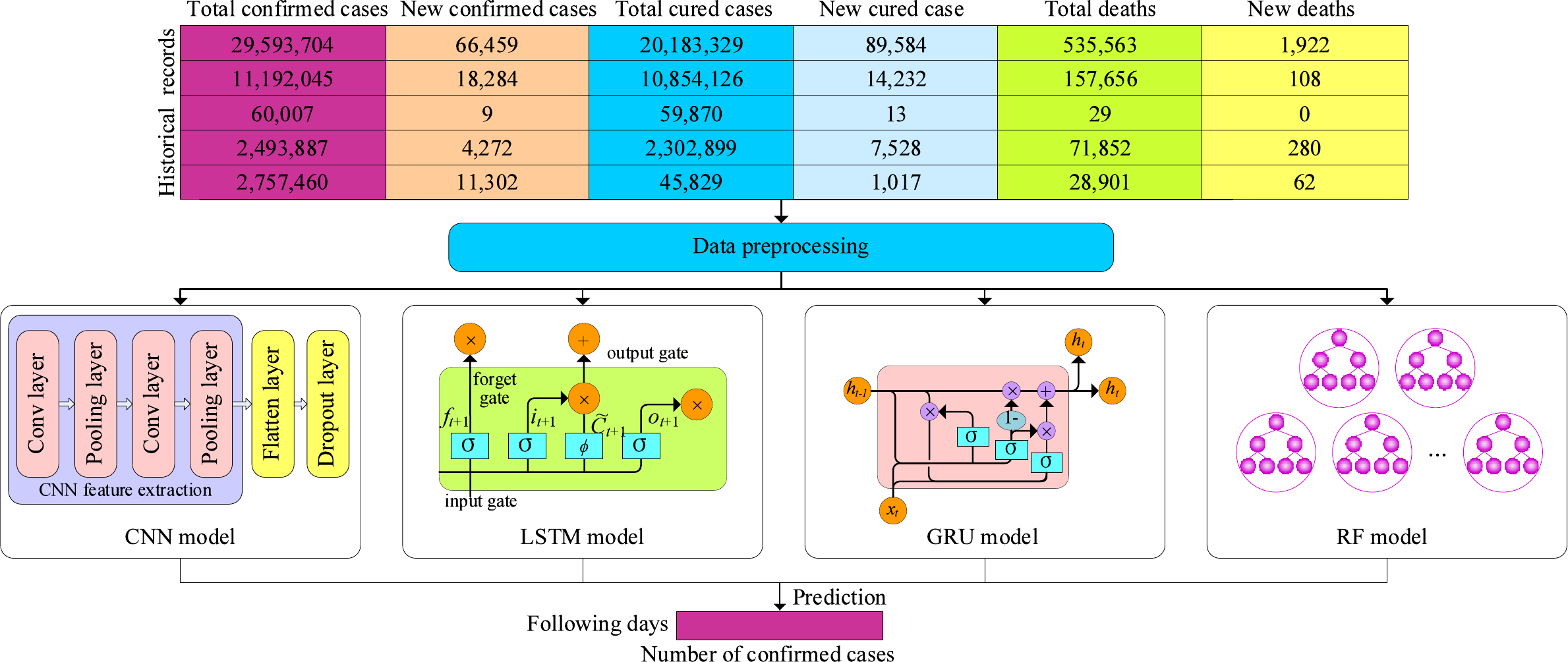}
 \caption{Representative AI architecture used for COVID-19 outbreak prediction.}
\label{fig11}
\end{figure}

Although the source of the COVID-19 epidemic has not yet been identified, it was first reported in Wuhan, China.
Therefore, the outbreak and spread of COVID-19 in China have received extensive attention.
In \cite{huang2020multiple}, Huang \emph{et al}. used 4 DL models (such as CNN, LSTM, GRU, and MLP) to train and predict COVID-19 cases from 7 severely epidemic cities in China.
The input of these DL models are the features of COVID-19 cases, including the number of confirmed cases, cured cases, and deaths.
Based on the input of the previous 5 days, each model can predict the number of COVID-19 cases in the following few days.
The representative AI architecture used for the COVID-19 outbreak prediction is shown in Fig. \ref{fig11}.

In \cite{hu2020artificial, hu2020evaluating}, Hu \emph{et al}. used AI methods such as MAE and clustering algorithms to predict the number of confirmed COVID-19 cases in different provinces and cities in China.
They divided 34 provinces and cities in China into 9 groups based on the prediction results, and further predicted the spread of COVID-19 between provinces and cities.
In \cite{yang2020modified}, Yang \emph{et al}. used the SEIR model \cite{keeling2008modeling} and the LSTM model to predict COVID-19 in China.
The population migration data and the latest COVID-19 epidemiological data from Baidu \cite{baidu2020data} were input into the SEIR model to derive the epidemic curves.
In addition, they used the 2003 SARS data to pre-train the LSTM model to predict COVID-19 in the following few days, in which epidemiological parameters (such as the transmission, incubation, recovery probability, and the number of deaths) were selected as input features.
Both SEIR and LSTM models predict the daily peak of infection in the first week of February will be 4,000.
In \cite{fong2020finding, fong2020composite}, Fong \emph{et al}. obtained early COVID-19 epidemiological data from NHC \cite{nhc2020data}.
Then, they used traditional time series data analysis methods (such as ARIMA, Exponential, and Holt-Winters), ML methods (such as KR, SVM, and DT), and AI methods (such as PNN) to analyze and predict future outbreaks.

In addition to China, the outbreak and spread of COVID-19 in other countries (including the United States, Italy, Spain, Iran, and Switzerland) have also received widespread attention.
In \cite{al2020optimization}, Ai \emph{et al}. proposed an improved ANFIS method \cite{jang1993anfis} to predict the number of COVID-19 cases.
The proposed system connects fuzzy logic and neural networks, and uses an enhanced flower pollination algorithm \cite{yang2012flower} for model parameter optimization and model training.
In \cite{rizk2020covid}, Rizk \emph{et al}. proposed an improved Multi-layer Feed-forward Neural Network (ISACL-MFNN) model, which uses an internal search algorithm (ISA) to optimize model parameters and a CL strategy to enhance ISA performance.
From the official COVID-19 data set reported by the WHO \cite{WHO2020data}, data from January 22, 2020, to April 3, 2020, in the United States, Italy, and Spain were collected to train the ISACL-MFNN model and to predict the confirmed cases within the next 10 days.
In \cite{giuliani2020modelling}, Giuliani \emph{et al}. collected the number of infected people in various provinces of Italy \cite{Italy2020data}, and used the EMTMGL model to simulate and predict the spatial and temporal distribution of COVID-19 infection in Italy.
They collected daily epidemic data and saved them in a time series data format, and then used LR and LSTM models to make predictions, thereby obtaining the outbreak and spread trend of COVID-19 in Iran.
In \cite{marini2020covid, marini2020enhancing}, Marini \emph{et al}. developed an agent-based AI platform, which accepts the entire Swiss population as input data to simulate and predict the spread of COVID-19 in Switzerland.
It simulates people's daily trajectories by calibrating micro-census data, and effectively predicts individual contacts and possible transmission routes.

Many studies have likewise focus on predicting the spread of COVID-19 worldwide.
They collected a large amount of travel data, mobile phone data, and social media data, and used AI methods to accurately predict the potential spread and transmission route of COVID-19.
In \cite{lai2020assessing}, Lai \emph{et al}. collected a large amount of travel and mobile phone data from \cite{worldpop2020data}, and constructed a corresponding model to predict the risk of COVID-19 transmission in different countries.
On this basis, they established an air travel network model between domestic cities and cities in other countries to predict risk cities at home and abroad.
In \cite{punn2020covid}, Punn \emph{et al}. used 2 ML models (e.g., SVR \cite{awad2015support} and PR \cite{de2019approximate}) and 3 DL regression models (e.g., DNN, LSTM \cite{greff2016lstm}, and RNN) to predict real-time COVID-19 cases.
In \cite{lampos2020tracking}, Lampos \emph{et al}. used an automatic crawling tool to obtain daily confirmed COVID-19 case data and related articles from online media, such as MediaCloud \cite{mediaCloud2020data}, Public Health England (PHE) \cite{govuk2020data}, and European Centre for Disease Prevention and Control (ECDC) \cite{ecdc2020data}.
They used the TL strategy to transfer the COVID-19 model of the disease-spreading country that are still in the early stage of the epidemic curve, achieving the epidemic prediction in the target country.
In addition, companies such as Microsoft Bing \cite{bing}, Google \cite{google2020data}, and Baidu \cite{baidu2020data} have aggregated multiple available data sources and developed a COVID-19 global tracking system to provide a visual tracking interface.

In addition to AI methods, various methods based on statistics and epidemiology are also used to predict the outbreak and spread of COVID-19.
In \cite{he2020temporal}, He \emph{et al}. collected the highest viral load in the pharyngeal swabs of 94 patients with confirmed COVID-19.
They fitted a generalized additive model with identification links and smooth spline curves to analyze their overall trend.
A gamma distribution was fitted to the transmission pair data to evaluate the serial interval distribution.
The results of statistical analysis indicated that the patients with confirmed COVID-19 have reached the peak of virus shedding before or during symptom onset, and some kinds of transmission may occur before the initial symptoms appear.
In \cite{wang2020tracking}, Wang \emph{et al}. determined a set of technical indicators that reflect the infection status of COVID-19, such as the number of infection cases in the hospital, daily infection rate, and daily cured rate.
Next, they proposed a calculation method based on statistical theory to quantify the iconic characteristics of each period and predict the turning point in the development of the epidemic.
In addition, numerous studies based on the Susceptible-Infected-Recovered (SIR) and SEIR models have studied the spread of COVID-19 from an epidemiological perspective.
Please see \cite{sameni2020mathematical, teles2020predicting, toda2020susceptible, caccavo2020chinese, siwiak2020single, zareie2020model, liu2020covid} for more information.

\subsection{Social Control}
When COVID-19 appeared, most countries in the world adopted different forms of social control, social alienation, school closures, and blockade measures to prevent the spread of the epidemic \cite{viner2020school}.
AI technologies have been widely used in epidemic control and social management, including individual temperature detection, video tracking, contact tracking, intelligent robots, etc.
Many countries have used smart devices equipped with AI to detect suspicious persons in public transportation places, such as airports and train stations.
For example, infrared cameras are used to scan high temperatures in a crowd, and different AI methods perform efficient analysis to detect whether an individual is wearing a mask in real time.
In addition, DL-based video tracking technology was used to detect and track suspicious COVID-19 patients in public places \cite{chen2019distributed}.
Moreover, at the entrances and exits of cities, the identity information of each passing person was collected.
Then, AI-based systems are used to efficiently query the travel history and trajectory of each passing individual to check whether they are from areas seriously affected by COVID-19 \cite{chen2020covid, tian2020investigation}.

AI technology is also used for contact tracking of patients with COVID-19 \cite{ keeling2020efficacy}.
For each patient with confirmed COVID-19, personal data (such as mobile phone positioning data, consumption records, and travel records) can be integrated to identify potential transmission trajectories \cite{ ge2020role}.
In addition, when people are in a state of social isolation, mobile phone positioning and AI frameworks can assist the government better understand the status of individuals \cite{rao2020identification}.
Intelligent robots are used to perform on-site disinfection and product transfer, and mobile phone positioning functions are used to detect and track the distribution and flow of personnel.
During the COVID-19 pandemic, Bluetooth technology was used in contact tracing APPs and deployed in many countries \cite{jeremy}.
Moreover, various countries/regions such as China, Canada, Singapore, Australia, France, Germany, and the United Kingdom, have deployed different contact tracing systems to effectively track public contacts and locate persons suspected of COVID-19 \cite{kleinman2020digital}.
For example, in China, a system of QR code was used to track the infected or potentially infected people \cite{kostka2020times}.
The Canadian government has published several Apps for COVID-19 contact tracing, such as the COVID Alert App, Canada COVID-19 self-assessment tool, and ArriveCAN \cite{canada}.
In Singapore, a real-time locating system (TraceTogether App) was developed to track public contacts during the COVID-19 pandemic \cite{ho2020use}.

Another group of studies focuses on the impact of various social control strategies on the spread of COVID-19.
In \cite{hellewell2020feasibility}, Hellewell \emph{et al}. proposed a random propagation model to analyze the success rate of different social control approaches to prevent the spread of COVID-19.
By quantifying parameters such as contact tracking and quarantined cases, the model analyzes the maximum number of cases tracked every week and measures the feasibility of public health work.
In \cite{kissler2020social}, Kissler \emph{et al}. established a mathematical model to assess the impact of interventions on the prevalence of COVID-19 in the United States, and recommended the use of intermittent grooming measures to maintain effective control of COVID-19.
Koo \emph{et al}. \cite{koo2020interventions} applied the population and personal behavior data of Singapore to the influenza epidemic simulation model and assessed different social isolation strategies for the dynamic transmission of COVID-19.
Similarly, Chang \emph{et al}. \cite{chang2020modelling} constructed a simulation model to evaluate the propagation and control of COVID-19 in Australia.
The model analyzes the propagation characteristics of COVID-19 and the impact of different control strategies on the propagation results.

\section{Data and Resources}
\label{section6}
The implementation and performance improvement of AI greatly depends on a large-scale available data and resources.
Therefore, we compiled available public resources that can be used for COVID-19 disease diagnosis, virology research, drug and vaccine development, and epidemic and transmission prediction.
Three types of data and resources are summarized, including medical images, biological data, and informatics resources.

\subsection{Medical Images}

We collected 17 groups of COVID-19 medical images, such as CXR and CT images, from researchers and organizations.
Among them, the CXR image data set published by Cohen \emph{et al}. \cite{cohen2020covid} is widely cited.
It is a collection of CXR images from multiple references.
In addition, many researchers uploaded CXR and CT images to Kaggle \cite{kaggle-xray, kaggle-xray2, mooney, kaggle2020data3, kaggle-xray3} for COVID-19 research.
Moreover, organizations such as the British Society of Thoracic Imaging (BSTI), Eurorad, and Radiopaedia also released online CXR and CT images.
Table \ref{table07} displays a detailed description of the medical image data resources of COVID-19.

\begin{table}[!ht]
\renewcommand{\arraystretch}{1.1}
\setlength{\abovecaptionskip}{4pt}
\setlength{\belowcaptionskip}{4pt}
\caption{Medical image data resources for COVID-19 research.}
\centering
\small
\label{table07}
\tabcolsep1pt
\begin{tabular}{L{2in} L{1.2in}  L{2in}}
\hline
Data sources           & Data type   &  Cited by Refs. \\
\hline
Zhao \cite{zhao2020covid} & CT images &  \cite{zhao2020covid} \\
Coronacases \cite{coronacases2020data}       & CT images  & - \\
Medical segmentation \cite{ms2020data}       & CT images & - \\
Cohen \cite{cohen2020covid}  & CXR images& \cite{hemdan2020covidx,wang2020covid,apostolopoulos2020covid,narin2020automatic,afshar2020covid,maghdid2020diagnosing,sethy2020detection,zhang2020covid}  \\
Wang \cite{wang2017hospital}  & CXR images & \cite{zhao2020covid,zhang2020covid}\\
COVIDx \cite{Chung}                 & CXR images   & \cite{wang2020covid} \\
Adrian \cite{Adrian}                & CXR images &   \cite{hemdan2020covidx} \\
COVID-Net \cite{wang2020covid}     & CXR images    &  \cite{wang2020covid} \\
Kermany \cite{kermany2018identifying} & CXR images & \cite{apostolopoulos2020covid} \\
Mendeley data \cite{Ali2020data} & CXR images & - \\
Kaggle \cite{kaggle-xray, kaggle-xray2, mooney, kaggle2020data3, kaggle-xray3}  & CXR and CT images & \cite{wang2020covid,apostolopoulos2020covid,narin2020automatic,afshar2020covid,sethy2020detection,maghdid2020diagnosing, Sarkar2020a} \\
BSTI \cite{BSTI}                   & CXR and CT images  & \cite{maghdid2020diagnosing} \\
SIRM \cite{sirm2020data}           & CXR and CT images  &  -  \\
Eurorad\cite{eurorad2020data}    & CXR and CT images   & - \\
Radiopaedia \cite{radiopaedia2020data}   & CXR and CT images & -\\
\hline
\end{tabular}
\end{table}

\subsection{Biological Data}
We collected 10 biological data resources, such as NCBI, Protein Data Bank (PDB), UniProt, Clarivate Analytics Integrity (CAI), Drug Target Common (DTC), and Virus-Host DB (VHDB), as shown in Table \ref{table08}.
These data resources provide abundant biological data resources, including gene sequences, proteins, drug molecules and compounds, and miRNA sequences.

\begin{table}[!ht]
\renewcommand{\arraystretch}{1.1}
\setlength{\abovecaptionskip}{4pt}
\setlength{\belowcaptionskip}{4pt}
\caption{Biological data sources for COVID-19 research.}
\centering
\small
\label{table08}
\tabcolsep1pt
\begin{tabular}{L{0.98in} L{1.2in} L{2.35in} L{0.8in}}
\hline
Data sources           & Data type   & Description  &  Cited by Refs. \\
\hline
NCBI \cite{ncbi2020data} & Genome sequences & Genome sequencing data of SARS-CoV-2 & \cite{randhawa2020machine, demirci2020computational, beck2020predicting, shin2019self, ong2020covid}\\
GISAID \cite{gisaid2020data} & Genome sequences & Bat Betacoronavirus RaTG13 &  \cite{randhawa2020machine} \\
VHDB \cite{mihara2016linking} & Genome sequences & Virus sequences & \cite{randhawa2020machine}\\
PDB \cite{pdb} & Proteins & 3D shapes of proteins, nucleic acids, and assemblies & \cite{ortega2020role, berman2003announcing} \\
UniProt \cite{bairoch2005universal} & Proteins &  SARS-CoV-2 protein entries and receptors & \cite{ong2020covid} \\
miRBase \cite{kozomara2019mirbase} & miRNA sequences & Human mature miRNA sequences &  \cite{demirci2020computational} \\
ZINC \cite{zinc2020data} & Drug compounds & drug compounds and molecules & \cite{kadioglu2020identification,moskal2020suggestions, hofmarcher2020large}\\
DTC \cite{tanoli2018drug} & Drug molecules &  Drug molecules for 3C-like proteases & \cite{beck2020predicting, shin2019self}\\
CAI \cite{cai2020data} & Drug discovery & Empowering knowledge-based drug discovery and development & \cite{zhavoronkov2020potential} \\
BindingDB \cite{liu2007bindingdb} & Amino-acid sequences & Amino-acid sequences of 3C-like proteases & \cite{beck2020predicting, shin2019self}\\
\hline
\end{tabular}
\end{table}

{\color[rgb]{0,0,1}
\subsection{Cough Sound Datasets}
Compared with medical images and biological data, cough sound data sets that can be used for COVID-19 analysis are very scarce.
In order to promote research in this direction, we collect 8 groups of human cough sound data sets, 5 of which contain cough sounds of patients with confirmed COVID-19, as shown in Table \ref{table10}.
We hope that these data sets will facilitate the diagnosis of COVID-19 through cough sounds.

\begin{table}[!ht]
\renewcommand{\arraystretch}{1.1}
\setlength{\abovecaptionskip}{4pt}
\setlength{\belowcaptionskip}{4pt}
\caption{Cough sound datasets for COVID-19 research.}
\centering
\small
\label{table10}
\tabcolsep1pt
\begin{tabular}{L{2in} L{1.6in} L{1.6in}}
\hline
Data sources           & Total number of records & Number of COVID-19 records   \\
\hline
COUGHVID \cite{zhao2020covid}     & 20,072 & 1,010 \\
Brown \cite{brown2020exploring}   & 9,986 & 235 \\
Virufy \cite{chaudhari2020virufy} & 3,349 & 517 \\
Coswara \cite{sharma2020coswara}  & 1,543  & 1,334 \\
Audio set \cite{gemmeke2017audio} & 871   & - \\
FREESOUND \cite{freesound}        & 736   & - \\
IIIT-CSSD \cite{singh2016iiit}    & 110   & - \\
Nococoda \cite{cohen2020novel}    & 73     & 73 \\
\hline
\end{tabular}
\end{table}
}

\subsection{Informatics Resources}
Informatics resources such as COVID-19 situation reports, dashboards, COVID-19 cases, and demographic data are gathered in Table \ref{table09}.
Among them, the World Health Organization (WHO), the National Health Commission of the People's Republic of China (NHC), and the Disease Control and Prevention (CDC) provide real-time COVID-19 reports.
Companies such as Baidu, Microsoft Bing, Google, and JHU CSSE provide online dashboards for COVID-19 spread tracking.
Organizations such as Govuk, Il Sole, and ECDC provide real-time COVID-19 cases in the UK, Italy, and Europe.

\begin{table}[!ht]
\renewcommand{\arraystretch}{1.1}
\setlength{\abovecaptionskip}{4pt}
\setlength{\belowcaptionskip}{4pt}
\caption{Informatics resources for COVID-19 research.}
\centering
\small
\label{table09}
\tabcolsep1pt
\begin{tabular}{L{1.05in} L{1.1in} L{2.35in} L{0.8in}}
\hline
Data sources           & Data type   & Description  &  Cited by Refs. \\
\hline
WHO \cite{WHO2020data} & Report & COVID-2019 situation reports & \cite{huang2020multiple, hu2020artificial, al2020optimization, rizk2020covid} \\
NHC \cite{nhc2020data} & Report &Real-time COVID-19 Report & \cite{fong2020finding, fong2020composite}\\
CDC \cite{cdc2020data}  & Report & Weekly U.S. influenza surveillance report & \cite{al2020optimization, lampos2020tracking}\\
GitHub \cite{github} & Code & Codes and data sets for COVID-19 study & \cite{Chung,zhao2020covid,ghddi2020data} \\
Baidu \cite{baidu2020data} & Dashboard & Dashboard for COVID-19 tracking & \cite{yang2020modified}\\
Bing \cite{bing} & Dashboard &  Dashboard for COVID-19 tracking & - \\
JHU CSSE \cite{dong2020interactive} & Dashboard & Dashboard for COVID-19 tracking & \cite{punn2020covid} \\
Govuk \cite{govuk2020data} &COVID-19 cases  & COVID-19 cases in the UK & \cite{lampos2020tracking} \\
ECDC \cite{ecdc2020data} & COVID-19 cases &COVID-19 cases worldwide & \cite{lampos2020tracking} \\
Il Sole \cite{Italy2020data} & COVID-19 cases & COVID-19 data sets in Italy & \cite{giuliani2020modelling}\\
Worldpop \cite{worldpop2020data} & Demographic data  & Spatial demographic and air travel data & \cite{lai2020assessing}\\
GHDDI \cite{ghddi2020data} & Community & Drug discovery community & \cite{hu2020prediction} \\
Humdata \cite{humdata}  & Community& community perceptions of COVID-19 & - \\
\hline
\end{tabular}
\end{table}

\section{Challenges and Future Directions}
\label{section7}
Although AI technologies have been used in fighting against COVID-19 and many studies have been published, we note that the application and contribution of AI in this work is still relatively limited.
We summarize the main challenges currently faced by AI against COVID-19 and provide the corresponding suggestions.

\subsection{Challenges}
At present, the application of AI in COVID-19 research mainly faces 5 challenges:
\begin{itemize}
\item Lack of available large-scale training data.
    Most AI methods rely on large-scale annotated training data, including medical images and various biological data.
    However, due to the rapid outbreak of COVID-19, there is insufficient data available for AI.
    In addition, annotating training samples is very time-consuming and requires professional medical personnel.
\item Data imbalance between positive and negative samples.
    There is a serious proportional imbalance between positive COVID-19 samples and negative samples (data resources without COVID-19).
    For data sets of medical images, biology samples, and cough sounds, negative samples are relatively easy to obtain from historical publications.
    However, there are limited positive COVID-19 samples can be collected, which further affects the accuracy and robustness of AI algorithms in COVID-19 diagnosis.
\item Massive noisy data and rumors.
    Relying on the developed mobile Internet and social media, massive noise information and fake news about COVID-19 have been published on various online media without rigorous review.
    However, AI algorithms seem to be powerless in judging and filtering the noise and erroneous data.
    This problem limits the accuracy and application of AI algorithms, especially in epidemic prediction and transmission analysis.
\item Limited knowledge in the intersection of computer science and medicine.
    Many AI scientists are from computer science, but the application of AI in the COVID-19 battle requires in-depth cooperation in computer science, medical imaging, bioinformatics, virology, and many other disciplines.
    Therefore, it is crucial to coordinate the cooperative work of researchers from different fields and integrate the knowledge of multiple subjects to jointly respond to COVID-19.
\item Data privacy and human rights protection.
    In the era of big data and AI, the cost of obtaining personal privacy data is very low.
    Faced with public health issues such as COVID-19, many governments hope to obtain various types of personal information, including mobile phone positioning data, personal travel trajectory data, and patient disease data.
    How to effectively protect personal privacy and human rights during information acquisition and AI-based processing is an issue worthy of discussion and attention.
\end{itemize}

\subsection{Future Directions}
In addition to the applications investigated in this survey, AI can also contribute to the COVID-19 battle from the following 11 potential directions:
\begin{enumerate}
\item  Non-contact disease detection.
    In CXR and CT image detection, the use of non-contact automatic image acquisition can effectively avoid the risk of infection between radiologists and patients during the COVID-19 pandemic.
    AI can be used for patient posture positioning, standard section acquisition of CXR and CT images, and movement of camera equipment.
\item  Few-shot learning and transfer learning.
    Focusing on the serious imbalance between the positive COVID-19 samples and negative samples, few-shot Learning, meta-learning, and transfer learning have the potential to be further applied in COVID-19 research \cite{cohen2020predicting, minaee2020deep}.
    After learning a large amount of data in a certain category, few-shot Learning algorithms can quickly detect a new category and obtain high accuracy by learning only a small number of samples.
    Transfer learning relaxes the assumption that training data must be independent and identically distributed (i.i.d) with test data, and effectively solve the problem of insufficient training data and imbalanced sample categories.
\item Remote video diagnosis.
     AI and NLP technologies can be used to develop remote video diagnosis systems and chat robot systems, and provide the public with COVID-19 disease consultation and preliminary diagnosis.
\item Patient prognosis management.
    In addition to long-term tracking and management of patients with COVID-19, AI technology (such as intelligent image and video analysis) can also be used to automatically monitor patient behavior during follow-up monitoring and prognostic management.
\item Biological research.
     In the field of biological research, AI can be used to discover the protein structures and features of viruses by accurately analyzing biomedical information, such as large-scale protein structures, gene sequences, and viral trajectories.
\item Drug and vaccine development.
      AI can be used not only to discover potential drugs and vaccines, but also to simulate the interaction between drugs and proteins and between vaccines and receptors, thereby predicting the potential responses of patients with COVID-19 with different constitutions to drugs and vaccines.
\item Identification and filtering of fake news.
      AI can be used to reduce and eliminate fake news and noise data on online social media platforms to provide reliable, correct, and scientific information about the COVID-19 pandemic.
\item Impact simulation and evaluation.
       Various simulation models can use AI to analyze the impact of different social control strategies on disease transmission.
       Then, they can be used to explore more effective and scientific disease prevention and social control methods.
\item Patient contact tracking.
      By constructing social relationship networks and knowledge graphs, AI can identify and track the trajectories of people in close contact with patients with COVID-19, so as to accurately predict and control the potential spread of the disease.
\item Intelligent robots.
       Intelligent robots are expected to be used in applications such as disinfection and cleaning in public places, product distribution, and patient care.
\item Intelligent Internet of Things.
       AI is expected to be combined with the Internet of Things to be deployed in customs, airports, railway stations, bus stations, and business centers.
       In this case, we can quickly identify suspicious COVID-19 virus and patients through intelligent monitoring of the environment and personnel.
\end{enumerate}

\section{Conclusions}
\label{section8}

In this survey, we investigated the main scope and contributions of AI in combating COVID-19.
Compared with the pandemic of SARS-CoV in 2003 and MERS-CoV in 2012, AI technologies have been successfully applied to almost every corner of the COVID-19 battle.
The application of AI in COVID-19 research can be summarized in four aspects, such as disease detection and diagnosis, virological research, drug and vaccine development, and epidemic and transmission prediction.
Among them, medical image analysis, drug discovery, and epidemic prediction are the main battlefields of AI against COVID-19.
We also summarized the currently available data and resources for AI-based COVID-19 research, including medical imaging data, biological data, and informatics resources.
Finally, we highlighted the main challenges and potential directions in this field.
This survey provides medical and AI researchers with a comprehensive view of the current and potential contributions of AI in combating COVID-19, with the goal of inspiring them to continue to maximize the advantages of AI and big data to combat this pandemic.


\bibliographystyle{abbrv}
\bibliography{sample-base}

\end{document}